\documentclass[journal,comsoc]{IEEEtran}

\usepackage{amsmath,amsfonts}
\usepackage{algorithmic}
\usepackage{algorithm}
\usepackage{array}
\usepackage{textcomp}
\usepackage{stfloats}
\usepackage{url}
\usepackage{verbatim}
\usepackage{graphicx}
\usepackage{cite}
\usepackage{hyperref}
\usepackage{balance}
\usepackage{enumitem}
\usepackage{wasysym}
\usepackage[weather]{ifsym}
\usepackage{textcomp}

\renewcommand{\texttt}[1]{%
	\begingroup
	\ttfamily
	\begingroup\lccode`~=`/\lowercase{\endgroup\def~}{/\discretionary{}{}{}}%
	\begingroup\lccode`~=`[\lowercase{\endgroup\def~}{[\discretionary{}{}{}}%
	\begingroup\lccode`~=`.\lowercase{\endgroup\def~}{.\discretionary{}{}{}}%
	\catcode`/=\active\catcode`[=\active\catcode`.=\active
	\scantokens{#1\noexpand}%
	\endgroup
}

\usepackage{tabularx}
\usepackage{booktabs}
\usepackage{makecell}
\usepackage{colortbl}
\usepackage{multirow}
\usepackage{array}
\usepackage{arydshln}
\newcommand*\dhline{\specialrule{0pt}{3pt}{0pt}\hdashline[.4pt/3pt]\specialrule{0pt}{0pt}{3pt}}
\newcommand{\dcline}[1]{\specialrule{0pt}{2pt}{0pt}\cdashline{#1}[.4pt/3pt]\specialrule{0pt}{0pt}{2pt}}

\usepackage{caption}
\usepackage{subcaption}
\usepackage{tikz}
\usetikzlibrary{shapes,arrows,arrows.meta,decorations,decorations.pathreplacing,calc,positioning,fit,shapes.arrows,math,backgrounds,fit,trees,spy,patterns,patterns.meta,colorbrewer}
\usepackage{pgfplots}
\usepgfplotslibrary{colorbrewer,groupplots}
\pgfplotsset{compat=1.18}
\usepackage{pgfplotstable}
\usetikzlibrary{external}
\usepgfplotslibrary{external}
\let\orgautoref\autoref
\renewcommand{\autoref}
{\def\sectionautorefname{Section}%
\def\subsectionautorefname{Section}%
\def\subsubsectionautorefname{Section}%
\orgautoref}

\usepackage{pifont}
\newcommand{\cmark}{\ding{51}}%
\newcommand{\xmark}{\ding{56}}%

\usepackage{wasysym}

\usepackage{xspace}
\newcommand{\etal}{\textit{et al.}}
\newcommand{\eg}{\textit{e.g.,}~}
\newcommand{\ie}{\textit{i.e.,}~}

\newcommand{\etc}{\textit{etc.}\xspace}
\newcommand{\one}{({\em i})\xspace}
\newcommand{\two}{({\em ii})\xspace}
\newcommand{\three}{({\em iii})\xspace}
\newcommand{\four}{({\em iv})\xspace}

\makeatletter
\renewcommand{\paragraph}[1]{\vspace*{0.03in}\noindent{\bf #1.}\hspace{0.25ex \@plus1ex \@minus.2ex}}
\makeatother

\begin{document}
	\urlstyle{tt}
	
	\title{How to Measure TLS, X.509 Certificates,\\ and Web PKI: A Tutorial and Brief Survey}
	
	\author{Pouyan Fotouhi Tehrani, Eric Osterweil, Thomas~C.~Schmidt, and~Matthias~W\"ahlisch%
		\thanks{Pouyan Fotouhi Tehrani is with TUD Dresden University of Technology, Germany (e-mail: pft@ieee.org).}%
		\thanks{Eric Osterweil is with George Mason University, USA (e-mail: eoster@gmu.edu).}%
		\thanks{Thomas C. Schmidt is with the Hamburg University of Applied Sciences~(HAW), Germany (e-mail: t.schmidt@haw-hamburg.de).}%
		\thanks{Matthias~W\"ahlisch is with TUD Dresden University of Technology, Germany. (e-mail: m.waehlisch@tu-dresden.de).}%
	}
		
	\maketitle

	\begin{abstract}
	Transport Layer Security (TLS) is the base for many Internet applications and services to achieve end-to-end security.
	In this paper, we provide guidance on how to measure TLS~deployments, including X.509 certificates and Web PKI.
	We introduce common data sources and tools, and systematically describe necessary steps to conduct sound measurements and data analysis.
	By surveying prior TLS~meaurement studies we find that diverging results are rather rooted in different setups instead of different deployments.
	To improve the situation, we identify common pitfalls and introduce a framework to describe TLS and Web PKI measurements.
	Where necessary, our insights are bolstered by a data-driven approach, in which we complement arguments by additional measurements.
	\end{abstract}
	

	\section{Introduction}
\label{sec:intro}
Many Internet~applications and services use Transport Layer Security (TLS) to enable authentication, confidentiality, and integrity end-to-end on top of an otherwise insecure transport layer.
Establishing a secure TLS~connection involves specific protocol handshakes based on X.509 certificates, which are part of the Web PKI system.
With the advent of QUIC, TLS~principles are integral part of the transport layer, making a clear understanding of different TLS deployments even more important.

\begin{figure}
    \scriptsize
    \centering
    \input{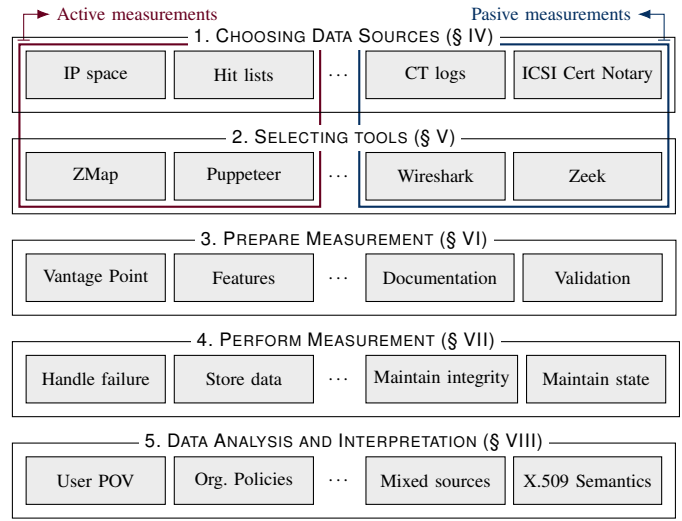}
    \caption{Stages of TLS and Web PKI measurement alongside selected aspects as discussed in this paper}
    \label{fig:intro}
\end{figure}

Due to the crucial role of TLS in securing Internet communication, the protocol itself as well as the different pieces such as X.509~certificates are popular subjects of Internet measurements.
Research questions range from understanding general usage of TLS and Web PKI, (\eg used algorithms and cryptographic material), deployment of related protocols (\eg DANE~\cite{RFC6698}), impact of security of incidents (\eg Heartbleed~\cite{Durumeric2014}), vulnerability to new attacks (\eg Confusion Attacks~\cite{DelignatLavaud2015}), adoption rate of TLS~versions (\eg TLS~1.3), to investigating nontechnical aspects such as market share of certification authorities.

Measuring and analysing TLS, X.509 certificates, and Web PKI is challenging due to \one application-specific implementations, \two inherent flexibility of TLS, and \three different measurement setups.
Applications may adapt TLS to their needs by defining further requirements and constraints.
This leads to TLS implementations with different mandatory features.
HTTP/2, for example, requires TLS to support Server Name Indication~(SNI) and prohibits use of specific cipher suites~\cite[\S 9.]{RFC7540}, while such constraints are not given for SMTPS~\cite{RFC3207}.
In terms of flexibility, TLS allows endpoints to negotiate parameters such as cryptographic algorithms during handshake.
Finally, differences between measurement setups make measurements that aim for answering the same questions not comparable and their findings inconsistent in some cases.

In this paper, we address challenges and pitfalls when measuring the TLS ecosystem.
To justify our arguments and illustrate pitfalls, we take a data-driven approach.
In addition to surveying prior work, we thus conduct our own measurements and provide three independent datasets.
Our main contributions are:

\begin{enumerate}
    \item A survey of prior TLS measurement research to illustrate common data collection techniques and measurement features.
    \item A framework to consistently exploring TLS and Web PKI deployments.
    \item An evaluation of common tools and data sources.
    \item Systematic guidance on how to prepare and perform measurements.
    \item A summary of most common pitfalls when interpreting measurement~data.
\end{enumerate}
We hope that these insights help to guide and improve future TLS measurements.

The remainder of this paper is structured as follows.
We start by providing a comprehensive picture of TLS and Web PKI~(\autoref{sec:bg}) and then introduce a simple framework to describe and compare measurements~(\autoref{sec:meas}).
In \autoref{sec:datasrc}, we review common data sources and discuss their merits.
\autoref{sec:tooling} comprises a list of measurement tools and their capabilities.
Subsequently, we discuss various aspects of preparing (\autoref{sec:prepare}) and performing (\autoref{sec:perform}) measurements followed by pitfalls when interpreting measurement data (\autoref{sec:interpret}).
We conclude this paper in \autoref{sec:conclusion}.
Related work that is the base for this paper is summarized in \autoref{tab:survey}.

	\section{Background}
\label{sec:bg}

\begin{figure*}
	\centering
	\input{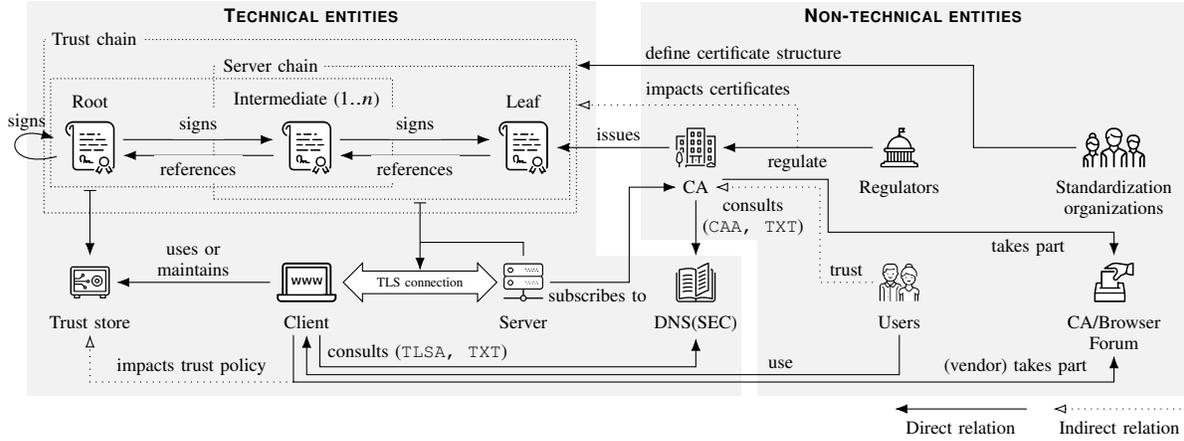}
	\caption{An overview of involved parties and components in TLS and Web PKI ecosystems}
	\label{fig:overview}
\end{figure*}

In this section we provide a brief overview of TLS, Web PKI, and DNS(SEC) and discuss related aspects.
\autoref{fig:overview} depicts technical components and non-technical entities involved in TLS and Web PKI ecosystems.

\subsection{Transport Layer Security}
\begin{figure}
	\scriptsize
	\centering
	\input{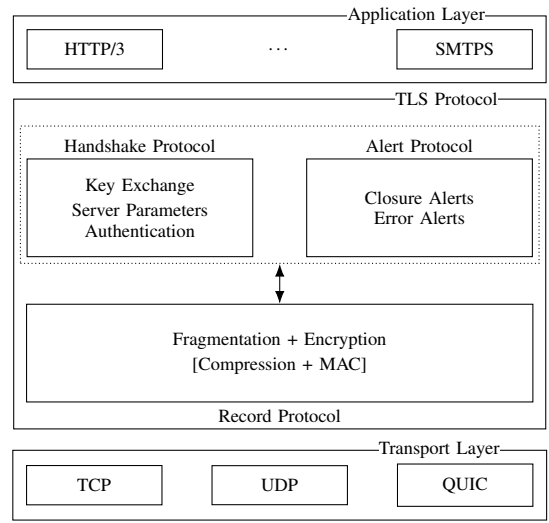}
	\caption{TLS Components: Handshake, Alert, and Record Protocols}
	\label{fig:tls-overview}
\end{figure}

TLS, the successor to Secure Sockets Layer (SSL), provides authenticated integrity and confidentiality over widely used transport layers of TCP, UDP, and QUIC~\cite{RFC8446,RFC9147,RFC9001}.
In this paper, we limit ourselves to TLS versions 1.2~\cite{RFC5246} and 1.3~\cite{RFC8446} as TLS versions 1.0 and 1.1 have already been deprecated~\cite{RFC8996}.

\paragraph{Components}
TLS protocol is split into two layers: the higher layer includes TLS-specific messages (\eg alerts and handshake messages), as well as applications data (\eg HTTP), and the lower layer, the \emph{record layer}, is responsible for fragmentation, encryption, and compression of higher-level messages before handing them to the transport layer.
An overview of TLS components and their relation to application and transport layer is given in \autoref{fig:tls-overview}.

\paragraph{Handshake}
To establish a secure channel, peers perform a handshake to determine security parameters, establish a shared key, and authenticate each other.
Server authentication in TLS~1.3 is mandatory, while in previous versions authentication is optional for both peers.
Authentication is commonly realized using X.509 certificates~\cite{RFC5280} or through pre-shared keys (PSK)~\cite{RFC4297,RFC8446}.

\paragraph{Extensions}
The functionality of TLS protocol can be extended using TLS extensions~\cite{RFC6066}\footnote{A list of registered extensions is made available by the Internet Assigned Numbers Authority (\emph{iana})~\cite{IANA2023}}.
Server Name Indication (SNI), and Stapled Online Certificate Status Protocol (OCSP) are among the prominent extensions.

\subsection{Web PKI and X.509}
The most widespread approach to authentication on the Internet is based on the Web PKI: an ecosystem of certification authorities (CA), applicants and subscribers (\eg website operators), and relying parties (\eg users).
A CA is a trusted third party (TTP) which validates the identity of an applicant and issues a corresponding X.509 certificate, \ie binds a public key to a subject.
The applicant then becomes a subscriber of the CA.
A relying party can subsequently authenticate the subscribers of its trusted CAs based on their presented certificates.

\begin{figure}
	\centering
	\input{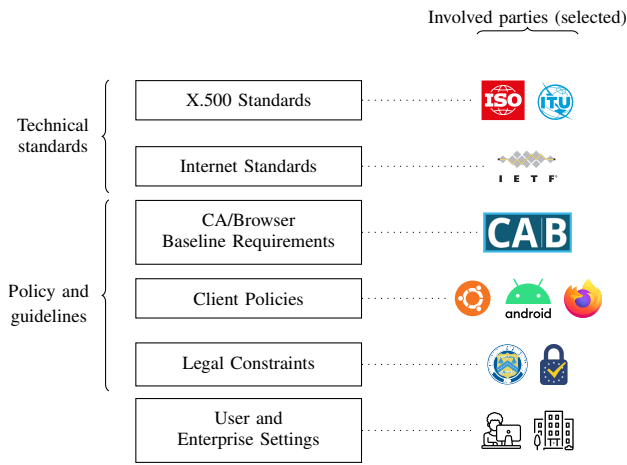}
	\caption{Technical and political aspects shaping X.509 Certificates in Web PKI}
	\label{fig:x509-hierarchy}
\end{figure}

\paragraph{Technical basis and governing entities}
The technical basis of X.509 certificates is standardized by the International Telecommunication Union (ITU) and International Organization for Standardization (ISO) as part of the X.500 standard series~\cite{iso9598-8}.
This is further specified and adapted by IETF for the Web PKI in RFC5280~\cite{RFC5280}.
A set of common guidelines and policies (Baseline Requirements) are also defined by the CA/Browser Forum~\cite{CABReq2.0}.
Additionally, each CA defines its own policies and procedures in Certificate Policy (CP) and Certification Practice Statement (CPS) documents.
And finally, individual organizations such as browser and operating system vendors impose their own requirements on X.509 certificates~(see for example Mozilla PKI Policy~\cite{MozillaPKI}).
\autoref{fig:x509-hierarchy} depicts this hierarchy.

\paragraph{Usage and semantics}
A certificate is used for authentication and identification through its binding of a public key to a subject.
Originally, a subject was meant to identify a natural or a legal person within a globally unique directory~(X.500~\cite{iso9594-1}), which never came into existence.
Alternative identities of the same subject from other namespaces (\ie domain name space) are included as Subject Alternative Names~(SAN).
Within the Web PKI ecosystem, the most basic form of an identity is a domain name.
Less common identities are IP and E-Mail addresses.

Before issuing a certificate, a CA validates the identity of an applicant regarding its ownership of included domain name(s) (Domain Validation, DV).
Additional validation procedures with higher identification assurances are Organization Validation (OV), Extended Validation (EV), and Individual Validation (IV)~\cite{CABReq2.0,CABEV}.
For DV validation, an applicant proves its control over respective domain name(s) through automated methods (\eg using ACME protocol~\cite{RFC8555}), or manually through e-mail or phone calls.
OV and IV certificates include additional identification information regarding respective organization or individual as defined in CA/B Baseline Requirements~\cite{CABReq2.0}.
EV certificates, originally introduced to make eCommerce more trustworthy, are subject to stricter requirements, audits, and lifecycle management~\cite{CABEV}.
Further validations types also exist for other purposes (\eg S/MIME authentication)~\cite{CABOID}.

\subsection{DNS and DNSSEC}
\label{subsec:bg-dns}
DNS is a hierarchical and distributed database with an eventual consistency model that maps domain names to Resource Records (RR) of various types, for example IPv4 and IPv6 addresses (\texttt{A} and \texttt{AAAA} records).
DNSSEC~\cite{RFC9364} is a set of security extensions that address a number of DNS security shortcomings (see RFC4033~\cite[\S 3]{RFC4033}).

In context of TLS and Web PKI, DNS(SEC) is utilized to counter various challenges.
As such, the holistic measurement of TLS and Web PKI deployment must also regard its relation to DNS.
In the following, we discuss three use cases of DNS(SEC) in context of TLS and Web PKI.

\paragraph{Removing ambiguity between CA and public keys}
CAs are generally not restricted in their certificate issuance~\cite[\S 4.2.1.10.]{RFC5280}, \ie a CA can issue a certificate for any arbitrary subject name (\eg domain name).
As such, a relying party cannot definitely determine if a given certificate was authorized by the entity denoted in the subject field or if it has been misissued.

The \textit{DNS-Based Authentication of Named Entities} (DANE) for TLS~\cite{RFC6698} addresses this ambiguity between public keys and CAs by allowing the owner of an identity (\eg a domain name or an e-mail address) to specify or characterize authorized certificates.
DANE relies on DNSSEC~\cite{RFC9364} and \texttt{TLSA} RRs.
An experimental TLS extension also enables embedding complete DNSSEC chains and \texttt{TLSA} records into the TLS handshake~\cite{RFC9102}.

DANE can be an effective remedy against rogue or compromised CAs as well as CAs used by intelligence services and governments~\cite{Soghoian2010} such as the root certificates generated and forced by Kazakhstan Government~\cite{KazakhtelecomJSC} or TrusCore CA which recently was disclosed to have ties to `contractors for U.S. intelligence agencies and law enforcement'~\cite{WP2022}.

\paragraph{Authorizing certifcate issuance}
Similar to relying party, CAs also need additional information to know if an applicant for an identity is actually authorized by identity owner (\eg a domain name owner) to avoid misissuance.
Domain name owners can use the CA authorization (\texttt{CAA}) resource record~\cite{RFC8659} to indicate which CAs are allowed to issue certificates for their domain names and also to report policy violation.

\paragraph{Upgrading opportunistic TLS}
While browsers commonly lack DANE support for HTTPS\cite{Zhu2015}, DANE has found popularity among SMTP servers~\cite{Lee2020} to address downgrade vulnerability of \texttt{STARTTLS}\cite{Alashwali2018}.

Mail server operators that are reluctant or unable to deploy DNSSEC, \eg Google, alternatively use \textit{SMTP MTA Strict Transport Security} (MTA-STS)~\cite{RFC8461}.
In this approach a \texttt{TXT} resource record under a well-known subdomain is used to signal TLS support by respective mail servers.
Related policies can then be fetched over HTTPS from a well-known URI~(see~RFC8615~\cite{RFC8615}).

	\section{Principles for a TLS Measurement Framework}
\label{sec:meas}

\begin{figure}
	\centering
	\footnotesize
	\input{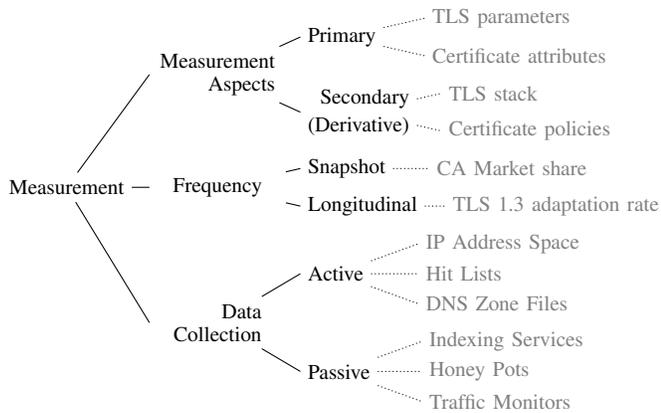}
	\caption{A taxonomy for TLS and Web PKI measurement with selected examples (in {\color{gray}gray})}
	\label{fig:meas-tax}
\end{figure}

The first step in measuring TLS and Web PKI is to decide on measurement \textbf{aspects} and choose a set of \textbf{features}.
The next step is to choose the measurement \textbf{frequency} with regards to the measurement goal, and finally to \textbf{collect data}.
In the following we discuss each aspect briefly and introduce a systematic approach to define and characterize TLS and Web PKI measurements.
Finally, we survey related work (cited in this paper) in context of our approach (\autoref{tab:survey}).

\subsection{Measurement Aspects and Features}
We differentiate between two types of measurement aspects.
Primary aspects can directly be measured and are quantifiable: X.509 public Key length, TLS session validity period, \etc are examples of this type.
Secondary or derivate ones can be inferred based on primary aspects or from raw data and might not be quantifiable: certificate policy statements, TLS stack in use by a server, and CA market share, for example.
Derivative aspects might be useful in interpreting data such as explaining irregular observations (see \autoref{sec:interpret}).

Here, we provide a set of primary aspects for TLS and Web PKI and limit our scope to study of TLS usage by Web servers, and X.509 certificates used for client or server authentication.
Note that some TLS and Web PKI aspects are associated with DNS(SEC) records and require collecting respective records.
We refer to related work~\cite{Osterweil2008,Osterweil2022} for methods of DNS and DNSSEC measurement.

\paragraph{TLS}
TLS can be measured regarding each of its components (see \autoref{sec:bg}) and the integration of the protocol as a whole in context of an application:

\begin{enumerate}[label={\textbf{T\theenumi}}]
	\item\label{itm:T1} Conformance to standards (\eg proper implementation of TLS state machine~\cite{Beurdouche2015,Ruiter2015,Levillain2021})
	\item\label{itm:T2} Handshake layer features (\eg supported extensions~\cite{RFC6066})
	\item\label{itm:T3} Record layer features (\eg susceptibility to known vulnerabilities)
	\item\label{itm:T4} Integration with higher level application layer (\eg TLS use by HTTP/2~\cite[\S 9.2.]{RFC7540})
\end{enumerate}

\paragraph{Web PKI}
As a complex ecosystem, Web PKI, can be studied regarding various aspects including its global structure, established trust stores, and CA certification policy and procedures.
X.509 certificates, at the core of Web PKI, can further be analyzed regarding the following properties:
\begin{enumerate}[label={\textbf{X\theenumi}}]
	\item\label{itm:X1} Conformance to existing standards (RFCs, CA/B baseline requirements, \etc~\cite{Durumeric2013,Brubaker2014,Nie2023,Quan2020})
	\item\label{itm:X2} Validity (in time and within a trust chain or trust store)
	\item\label{itm:X3} Extensions and expected usage (\eg validation method and key usage flags)
	\item\label{itm:X4} Revocation status (OCSP and CRL)
	\item\label{itm:X5} Relation to relevant DNS records (\eg \texttt{CAA} and \texttt{TLSA})
\end{enumerate}

\subsection{Frequency}
In specific cases, it is desired or required to perform multiple measurements.
If we are interested in observing temporal developments and discovering causal relations, \eg in response to introduction of a new technology such as QUIC, we need to measure the whole sample set multiple times over a period of time.
Such longitudinal measurements can be considered as regular repetition of snapshot measurements.

\paragraph{Snapshot}
This type of measurement is based on probing a sample set on a set of features at a given point in time.
Such measurements help to understand state of affairs at a single point in time.
Shobiri~\etal~\cite{Shobiri2023}, for example, investigates if CDN provides observe security hygiene when communicating to backend servers by monitoring incoming connections from CDN edge servers.

\paragraph{Longitudinal}
When the same set of features for a constant sample set is probed over a period of time, we speak of longitudinal measurements.
In contrast to snapshots, longitudinal data are suitable for understanding causal relations.
Probing same targets over time enables capturing temporal changes, \eg in response to an incident.
Durumeric~\etal~\cite{Durumeric2014}, for example, analyze (among others) the patch rate of HTTPS servers vulnerable to Heartbleed by probing Alexa Top 1M sites and a sample set of  IPv4 address space every 8 hours for a period of ca 3 months.
Similarly, Zhang~\etal~\cite{Zhang2014} study revocation and reissuance of certificates for vulnerable hosts at the wake of the Heartbleed.

\subsection{Data Collection}
Raw data can be collected by intercepting network (passive) or through initiation of purposeful transactions (active).

\paragraph{Passive}
In passive measurements, data in transit, \ie passing through the network, is captured without further modification.
This presupposes access to special network nodes, \eg routers.
Passive measurements are useful for studying traits that cannot actively be measured, \eg actual cipher suites used by users.

In context of TLS, the main challenge of passive monitoring regards relevant data that might be encrypted and cannot be dissected, \eg after the handshake concludes or upon TLS session resumptions.
To address this issue, data can be collected directly at endpoints instead of in transit.
Oakes~\etal~\cite{Oakes2019}, for example, collect data from a web monitoring tool installed directly on user devices.

\paragraph{Active}
In active measurements purposefully generated or manipulated data packets are exchanged with hosts.
Both request and response data are then recorded for further analysis.
Active measurement is suitable to investigate the behavior of a counterpart in presence of a concrete condition or data packet.
Hebrok~\etal~\cite{Hebrok2023}, for example, analyze security of TLS sessions by generating special handshake messages and testing these against TLS servers.

\newcommand{\cdash}{\multicolumn{1}{c}{--}}
\begin{table*}
	\scriptsize
	\addtolength{\tabcolsep}{-0.4em}

	\caption{Brief survey of related work that has been cited in this work}
	\label{tab:survey}
	\begin{tabularx}{\textwidth}{c p{2.5cm} p{1cm} p{4.5cm} c X X}
		\toprule
		& & & & \multicolumn{2}{c}{Measurement Features} & \\
		\cmidrule{5-6}
		Year & Authors & Context & \multicolumn{1}{c}{Data collection\textsuperscript{$\dagger \ddagger$}} & Fundamental & \multicolumn{1}{c}{Derivative} & \multicolumn{1}{c}{Frequency} \\
		\midrule

		\multirow{4}{*}{2011} & \multirow{4}{*}{Holz~\etal~\cite{Holz2011}} & \multirow{4}{*}{generic} & \textcircled{A} TLS from DE over Alexa 1M & \multirow{4}{*}{\ref{itm:T2}~\ref{itm:X1}~\ref{itm:X2}~\ref{itm:X3}} & \multirow{4}{*}{CA market share} & 2009-11 -- 2011-04\textsuperscript{\text{\Lightning}} \\
		& & & \textcircled{A} TLS outside DE over Alexa 1M & & & 2011-04 \\
		& & & \textcircled{P} TLS research net traffic from DE & & & 2010-09 and 2011-04 \\
		& & & \textcircled{A} TLS from EFF over IPv4 & & & 2010-03 -- 2011-06 \\
		\midrule
		
		2013 & Durumeric~\etal~\cite{Durumeric2013a} & generic & \textcircled{A} X.509 over IPv4 & \ref{itm:X1}~\ref{itm:X2}~\ref{itm:X3}~\ref{itm:X4} & CA ownership and market share & 2012-06-06 -- 2013-08-04 \\
		\midrule

		2014 & Liang~\etal~\cite{Liang2014} & CDN & \textcircled{A} DNS, TLS, and HTTP over Alexa 1M & \ref{itm:X1}~\ref{itm:X2}~\ref{itm:X4} & Shared certificates in CDNs & \emph{undefined} \\
		\midrule

		\multirow{10}{*}{2014} & \multirow{10}{*}{Durumeric~\etal~\cite{Durumeric2014}} & \multirow{10}{*}{Heartbleed} & \textcircled{A} TLS over Alexa 1M & \multirow{2}{*}{\ref{itm:T1}~\ref{itm:T2}~\ref{itm:T3}~\ref{itm:X4}} & \multirow{2}{4.5cm}{· Vulnerable devices/software \\ · Patching behavior} & \multirow{2}{*}{2014-04 and 2014-05} \\
		& & & \textcircled{A} TLS over 1\% of IPv4 & & & \\
		\cmidrule{4-7}
		& & & \textcircled{A} TLS over IPv4 & \ref{itm:T2} & False negative rate & 2014-04-24, 2014-05-01\\
		\cmidrule{4-7}
		& & & \textcircled{P} TLS from ICSI over research net & \multirow{2}{*}{\ref{itm:T2}} & Pre-disclosure patching & 2014-04 \\
		& & & \textcircled{A} TLS from Uni Michigan over IPv4 & & Certificate replacement\textsuperscript{1} & 2014-04 -- 2014-05 \\
		\cmidrule{4-7}
		& & & \textcircled{P} TLS from LBNL net & \multirow{3}{*}{\ref{itm:T2}} & \multirow{3}{*}{Attack activity\textsuperscript{1}} & \makecell[l]{2012-02 -- 2012-03, \\ 2013-02 -- 2013-03, \\ 2014-01 -- 2014-04}\\
		& & & \textcircled{P} TLS from NERSC net & & & 2014-02 -- 2014-04 \\
		& & & \textcircled{P} TLS from private honeypot & &  & 2013-11 -- 2014-04 \\
		\midrule

		\multirow{5}{*}{2014} & \multirow{5}{*}{Zhang~\etal~\cite{Zhang2014}} & \multirow{5}{*}{Heartbleed} & \textcircled{A} X.509 from Rapid7 over IPv4 & \multirow{3}{*}{\ref{itm:T2}~\ref{itm:X2}~\ref{itm:X3}~\ref{itm:X4}} & \multirow{3}{4.5cm}{· Certificate replacement \\ · Time diff revocation and\\ ~~replacement} & 2013-10-30 -- 2014-04-28 \\
		& & & \textcircled{A} CRL endpoints of collected X.509 & & & 2014-05-06 \\
		& & & \textcircled{A} TLS over Alexa 1M & & & \emph{undefined} \\
		\cmidrule{4-7}
		& & & \textcircled{A} TLS over Alexa 1M (from~\cite{Durumeric2014}) & \multirow{2}{*}{\ref{itm:T2}} & \multirow{2}{*}{False negative rate} & 2014-04-09 \\
		& & & \textcircled{A} TLS over Alexa 1M & & & 2014-04-28 \\
		\midrule

		\multirow{3}{*}{2015} & \multirow{3}{*}{\makecell[tl]{Delignat-Lavaud\\ and Bhargavan~\cite{DelignatLavaud2015}}} & \multirow{3}{*}{\makecell[lt]{Confusion\\attacks}} & \makecell[tl]{\textcircled{A} X.509 from EFF and\\ Alexa 1M (from~\cite{Abadi2013,DelignatLavaud2014})} & \ref{itm:X3} & Shared certificates & 2013-07-31 \\
		\cmidrule{4-7}
		& & & \textcircled{A} HTTPS over Alexa 10k & \ref{itm:T4} & Cross-protocol redirection & \emph{undefined} \\
		\midrule

		2015 & Zhu~\etal~\cite{Zhu2015} & DANE & \textcircled{A} TLS over ca 1k domains w/ TLSA & \ref{itm:X5} & \texttt{TLSA} RR size and IP fragmentation & 2014-07-14--2014-12-03\textsuperscript{\text{\Lightning}}\\
		\midrule

		\multirow{4}{*}{2015} & \multirow{4}{*}{Liu~\etal~\cite{Liu2015}} & \multirow{4}{*}{generic} & \textcircled{A} X.509 from Rapid7 over IPv4 & \multirow{3}{*}{\ref{itm:X4}} & \multirow{3}{4.5cm}{· CA behavior on revocation \\ · CRL sizes and IP fragmentation} & 2013-10-31 -- 2015-03-30 \\
		& & & \textcircled{A} CRL endpoints of collected X.509 & & & 2014-10-02 -- 2015-03-31 \\
		& & & \textcircled{A} OCSP endpoints of collected X.509 & & & 2015-03-31 \\
		\cmidrule{4-7}
		& & & \textcircled{A} TLS from Uni Michigan over IPv4 & \ref{itm:T2}~\ref{itm:X4} & Stapled OCSP support & 2015-03-28 \\
		\midrule

		\multirow{2}{*}{2016} & \multirow{2}{*}{Chung~\etal~\cite{Chung2016}} & \multirow{2}{*}{generic} & \textcircled{A} X.509 from Uni Michigan over IPv4 & \multirow{2}{*}{\ref{itm:X1}\ref{itm:X2}\ref{itm:X4}} & \multirow{2}{4.5cm}{· CA and Device diversity\\· Certificate to device map\\ } & 2012-06-10--2014-01-29\textsuperscript{\text{\Lightning}} \\
		& & & \textcircled{A} X.509 from Rapid7 over IPv4 & & & 2013-10-30--2015-03-30 \\
		\midrule

		\multirow{2}{*}{2016} & \multirow{2}{*}{Holz~\etal~\cite{Holz2016}} & \multirow{2}{1cm}{E-Mail and Chat} & \textcircled{A} TLS over IPv4 & \multirow{2}{*}{\ref{itm:T4}~\ref{itm:X2}} & \multirow{2}{4.5cm}{Certificate sharing} & 2015-06-09--2015-08-04\textsuperscript{\text{\Lightning}} \\
		& & & \textcircled{P} TLS from UC Berkley net & & & 2015-07-29--2015-08-06 \\
		\midrule

		2016 & \makecell[l]{Springall, Durumeric,\\ and Halderman~\cite{Springall2016}} & TLS resumption & \textcircled{A} TLS over Alexa 1M & \ref{itm:X5} & \texttt{TLSA} RR size and IP fragmentation & 2014-07-14--2014-12-03\textsuperscript{\text{\Lightning}}\\
		\midrule

		\multirow{4}{*}{2016} & \multirow{4}{*}{Cangialosi~\etal~\cite{Cangialosi2016}} & \multirow{4}{1cm}{Key sharing} & \textcircled{A} X.509 from Rapid7 over IPv4 & \ref{itm:X3} & \makecell[lt]{· Domain ownership \\ · Certificate sharing} & 2013-10-30--2015-03-30 \\ 
		\cmidrule{4-7}
		& & & \textcircled{A} X.509 from Zhang~\cite{Zhang2014} (see above) & \ref{itm:X3}~\ref{itm:X4} & revocation of self-hosted vs managed & \emph{undefined} \\
		\midrule

		\multirow{8}{*}{2016} & \multirow{8}{*}{VanderSloot~\etal~\cite{VanderSloot2016}} & \multirow{8}{1cm}{Source coverage} & \textcircled{A} X.509 from Censys & \multirow{8}{*}{\ref{itm:X2}} & \multirow{8}{*}{--} & \multirow{8}{*}{2016-08-29--2016-09-08}\\
		& & & \textcircled{A} TLS over IPv4 & & & \\
		& & & \textcircled{A} TLS over Alexa 1M & & & \\
		& & & \textcircled{A} X.509 over CT logs & & & \\
		& & & \textcircled{A} X.509 over domains in CT logs & & & \\
		& & & \textcircled{A} TLS over \texttt{.com}, \texttt{.net}, \texttt{.org} & & & \\
		& & & \textcircled{A} TLS over domains in Common Crawl & & & \\
		& & & \textcircled{P} X.509 ICSI over research net & & & \\
		\midrule

		2017 & Ouvrier~\etal~\cite{Ouvrier2017} & generic & \textcircled{P} TLS from Uni Calgary net & \ref{itm:T2}~\ref{itm:T3}~\ref{itm:T4}~\ref{itm:X1}~\ref{itm:X3} & CA and browser market share & 2015-10-11--2015-10-17\\
		\midrule

		2017 & Amann~\etal~\cite{Amann2017} & generic & \textcircled{A} TLS over 193M domain names (from TUM and Uni Sydney) & \ref{itm:T2} \ref{itm:T4}& \multirow{2}{4.5cm}{· Prevalence of SCT \\ · Usage of HSTS and HPKP}  & \emph{undefined}\\
		\cmidrule{4-7}
		& & & \textcircled{P} TLS from UC Berkley net & & Prevalence of SCT &  \emph{undefined}\\
		\dcline{4-7}
		& & & \textcircled{P} TLS from TUM net & \multirow{2}{*}{\ref{itm:T2} \ref{itm:X2} \ref{itm:X3} \ref{itm:X4}} & \multirow{2}{*}{[Validating measurement above]} & \multirow{2}{*}{\emph{undefined}} \\
		& & & \textcircled{P} TLS from Uni Sydney & & & \\
		\cmidrule{4-7}
		& & & \textcircled{P} TLS from ICSI over research net & & TLS version evolution & 2012-04-01--2017-09-01\\
		\midrule
		
		\multirow{2}{*}{2018} & \multirow{2}{*}{Scheitle~\etal~\cite{Scheitle2018b}} & \multirow{2}{*}{CT Logs} & \textcircled{P} TLS from UC Berkley net &  \multirow{2}{*}{\ref{itm:T2} \ref{itm:X2} \ref{itm:X3} \ref{itm:X4}} & \multirow{2}{4.5cm}{· Prevalence of SCT \\ · Info leakage over CT logs} & 2017-04-26--2018-05-23\\
		& & & \textcircled{A} TLS over 423M domain names & & & 2018-05-18 \\
		\midrule
	\end{tabularx}
\end{table*}

\begin{table*}
	\scriptsize
	\addtolength{\tabcolsep}{-0.4em}

	\begin{tabularx}{\textwidth}{c p{2.5cm} p{1cm} p{4.5cm} c X X}
		2018 & Sy~\etal~\cite{Sy2018} & TLS Sessions & \textcircled{A} TLS over Alexa 1M & \ref{itm:T2} \ref{itm:T3} & Session sharing & 2018-04--2018-05 \\
		\midrule

		\multirow{2}{*}{2018} & \multirow{2}{*}{Kotzias~\etal~\cite{Kotzias2018}} & \multirow{2}{*}{generic} & \textcircled{P} TLS from ICSI SSL Notary & \multirow{2}{*}{\ref{itm:T2}} & \multirow{2}{4.5cm}{· TLS fingerprinting \\ · Vulnerability analysis} & 2012-02--2018-03 \\
		& & & \textcircled{A} TLS over IPv4 and Alexa 1M (Censys) & & & 2015-08--2018-05 \\
		\midrule

		\multirow{2}{*}{2018} & \multirow{2}{*}{Chung~\etal~\cite{Chung2018}} & \multirow{2}{*}{OCSP} & \textcircled{A} X.509 from Censys & \multirow{2}{*}{\ref{itm:T2} \ref{itm:X2} \ref{itm:X4}} & \multirow{2}{*}{OCSP responder availability} & \multirow{2}{*}{\emph{undefined}} \\
		& & & \textcircled{A} TLS over Alexa 1M & & & \\
		\midrule

		2019 & Oakes~\etal~\cite{Oakes2019} & generic & \textcircled{P} X.509 from 2M residential clients & \ref{itm:X2} \ref{itm:X3} & CA market share & 2017-07-12--2018-01-12\\
		\midrule

		\multirow{4}{*}{2019} & \multirow{4}{*}{Aas~\etal~\cite{Aas2019}} & \multirow{4}{1cm}{Let's Encrypt} & \textcircled{A} X.509 from Censys & \multirow{4}{*}{\ref{itm:T2} \ref{itm:X2}} & \multirow{4}{4.5cm}{· Prevalence among top sites \\ · CA market share} & \multirow{4}{*}{\emph{undefined}}\\
		& & & \textcircled{A} X.509 from CT logs & & & \\
		& & & \textcircled{A} TLS over names from CT logs & & & \\
		& & & \textcircled{P} Firefox client telemetry & & & \\
		\midrule

		2019 & Holz~\etal~\cite{Holz2019} & TLS~1.3 & \textcircled{A} TLS over Alexa 1M + \texttt{com}/ \texttt{net}/\texttt{org} & \ref{itm:T2} & Prevalence among TLDs and hosting services & 2019-05-\{1,3,4,5\}\\
		\cmidrule{4-7}
		& & & \textcircled{P} TLS from ICSI SSL Notary & \ref{itm:T2} \ref{itm:T3} & \multirow{2}{*}{TLS version prevalence} & 2012-03-01--2019-09-01 \\
		& & & \textcircled{P} TLS from an Australian uni net & \ref{itm:T2} & & 2019-05-09--2019-05-13\\
		\cmidrule{4-7}
		& & & \textcircled{P} TLS from Lumen Privacy Monitor & \ref{itm:T2} & Client/Server support & 2016-12-01--2019-04-01\\
		\midrule

		2020 & Lee~\etal~\cite{Lee2020} & DANE & \textcircled{A} TLS over DANE-enabled servers & \ref{itm:T2} \ref{itm:T4} \ref{itm:X2} & -- & 2019-07-11--2019-10-31\\
		\midrule

		2020 & Zhang~\etal~\cite{Zhang2020} & Confusion Attacks & \textcircled{A} TLS over domains from CT logs & \ref{itm:X3} & Shared certificates and context confusion attacks & \emph{undefined}\\
		\midrule

		2020 & Wan~\etal~\cite{Wan2020} & Source coverage & \textcircled{A} TLS over IPv4 & -- & -- & 2019-10--2019-12\textsuperscript{\text{\Lightning}}\\
		\midrule

		2020 & Raman~\etal~\cite{Raman2020} & Censorship & \textcircled{A} TLS to vantage points in Kazakhstan & \multirow{2}{*}{\ref{itm:T2} \ref{itm:X2}} & Interception triggers & 2019-07-20 \\
		& & & \textcircled{A} TLS to web servers in Kazakhstan & & Interception detection & 2019-07-\{22,23\} \\
		\midrule

		2021 & Lee~\etal~\cite{Lee2021} & Security & \textcircled{A} TLS over names from Rapid7 & \ref{itm:T2} \ref{itm:T3} & Location-dependent security & \emph{undefined}\\
		\midrule

		2021 & Brinkmann~\etal~\cite{Brinkmann2021} & Confusion Attacks & \textcircled{A} TLS over IPv4 & \ref{itm:T2} \ref{itm:X2} & Vulnerability to Cross-Protocol Attacks & \emph{undefined}\\
		\midrule

		2021 & Tatang, Flume, and Holz~\cite{Tatang2021} & MTA-STS & \textcircled{A} TLS over Tranco 1M & \ref{itm:T2} \ref{itm:X2} & MTA-STS prevalence & 2019-06-26--2019-10-13\textsuperscript{\text{\lightning}} \\
		\midrule

		2021 & Zhang~\etal~\cite{Zhang2021} & Root certs & \textcircled{P} TLS over \emph{360 Secure Browser} & \ref{itm:X2} & Evaluation of \emph{hidden roots} & 2020-02-01--2020-06-30 \\
		\midrule

		2021 & Ma~\etal~\cite{Ma2021a} & CA & \textcircled{P} X.509 from CT logs & \ref{itm:X3} & CA operators landscape & 2020-07-01 \\
		\midrule

		\multirow{2}{*}{2022} & \multirow{2}{*}{Nawrocki~\etal~\cite{Nawrocki2022}} & QUIC & \textcircled{A} TLS over Tranco 1M (TCP) & \multirow{2}{*}{\ref{itm:T4} \ref{itm:X3}} & \multirow{2}{*}{Impact of X.509 certs on QUIC} & \multirow{2}{*}{\emph{undefined}} \\
		& & & \textcircled{A} TLS over Tranco 1M (QUIC) & & & \\
		\midrule
		
		2022 & Lee~\etal~\cite{Lee2022} & DANE & \textcircled{A} TLS over \texttt{com} / \texttt{net} / \texttt{org} / \texttt{se} & \ref{itm:X2} \ref{itm:X3} \ref{itm:X5} & Correctness of key rollovers & 2019-07-13--2021-02-12\\
		\midrule

		\multirow{2}{*}{2023} & \multirow{2}{*}{Farhan, and Chung~\cite{Farhan2023}} & \multirow{2}{*}{generic} & \textcircled{A} X.509 over IPv4 from Rapid7 & \multirow{2}{*}{\ref{itm:X2}} & \multirow{2}{*}{CA market share evolution} & 2013-09--2021-12\\
		& & & \textcircled{P} X.509 from CT logs & & & 2013--2021-02\\
		\bottomrule

		\multicolumn{7}{l}{\shortstack[l]{
			\textsuperscript{$\dagger$} \textcircled{A}: active measurement, \textcircled{P}: passive measurement
			\textsuperscript{$\dagger$} includes only TLS and X.509 datasets
			\textsuperscript{\text{\Lightning}} Irregular measurement\\
			\textsuperscript{1} In combination with active and ICSI scans of the same study}}
	\end{tabularx}
\end{table*}

	\section{Data sources}
\label{sec:datasrc}

Here, we introduce various sources for data collection and discuss their merits: subsections \ref{subsec:src-ip} to \ref{subsec:src-dnszone} regard active measurements and the remainder are related to passive measurements.
In \autoref{sec:prepare}, we dig deeper into the ramifications of choosing specific data sources for measurements, and in \autoref{sec:interpret} we discuss pitfalls when interpreting data from specific sources.

\subsection{IP Address Space Scanning}
\label{subsec:src-ip}
a straightforward and commonly used approach to TLS and Web PKI measurement is first to scan the IP address space (or a selected subnet) for ports reserved for TLS-enabled protocols~(\eg 443 for HTTPS) and then to establish a TLS connection with matching hosts.
It should, however, be noted that since presence of an open port on its own is not a guarantee that respective protocol is also available at that endpoint~\cite{Durumeric2013,Holz2016}, false positives need then be discarded during the actual measurement.
In \autoref{sec:tooling}, we discuss tools that address both IP scanning and service detection.

A disadvantage of IP scanning is the lack of intrinsic ordering among IP addresses, \eg in terms of popularity or usage, so that at first glance the IP address of a home router (possibly accessed by a single individual) has the same weight as an IP address of a critical or high-traffic server, \eg \texttt{www.google.com}.
Additionally, detecting cases of opportunistic TLS, \eg \texttt{STARTTLS} over SMTP, require scanners to perform protocol handshakes to determine if a security upgrade is supported.
And finally, virtual hosts on the same server, \ie behind the same IP address, cannot individually be addressed in this approach, \eg using SNI extension~\cite{VanderSloot2016} (see \autoref{sec:bg}).

\begin{figure}
    \scriptsize
    \input{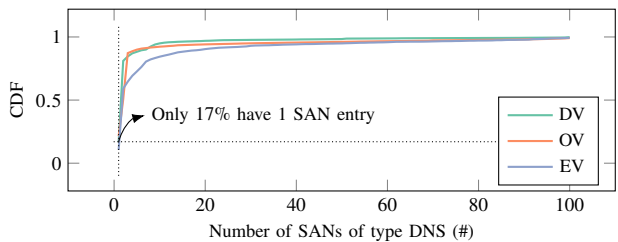}
    \caption{Distribution of number of DNS SANs (capped at 100) in X.509  certificates by validation type (Domain, Organization, or Extended Validation) from our Tranco dataset}
    \label{fig:san-dist}
\end{figure}

Research~\cite{Durumeric2013,Holz2011} shows how IP scanning can skew measurement results and result in an incomplete picture of TLS and Web PKI landscape, partly due to virtual hosting and required usage of Server Name Indication (SNI).
Although SNI is an optional TLS extension~\cite[\S 3]{RFC6066}, its use is mandated by some higher-level applications protocols.
HTTP/2~\cite[\S 9.2.]{RFC7540} and HTTP/3~\cite[\S 9.3.]{RFC9114}, for example, require clients to use SNI for the TLS connection if server is identified by a domain name.
Respectively, web servers behavior vary depending on existence of SNI.
Cloudflare, for example, aborts the TLS handshake (alert 40) if no SNI is provided, whereas Firebase (Google) and Wordpress return wildcard certificates (\texttt{*.firebaseapp.com} and \texttt{*.wordpress.com}) that would match the free subdomains allocated for their customers.
The possibility of encountering such cases can be elaborated by \autoref{fig:san-dist} which depicts the distribution of the number of DNS SAN in certificates from our \texttt{TRANCO} dataset (see \autoref{sec:appendix}).
We observe that only about $17\%$ of all certificates enlist a single SAN entry and among the rest about $16\%$ contain at least two unique eSLDs signalling that the certificate might be shared among two different subjects but served under the same IP.

\subsection{Hit Lists}
\begin{table}
	\scriptsize
	\caption{Comparison of various hit lists}
	\label{tab:hitlists}
	\begin{tabularx}{\columnwidth}{X l c c r}
		\toprule
		Name & Source & Scoped & Ordered & Entries \\
		\midrule
		Tranco & List Aggregation & \xmark & \cmark & $\approx3.6M$ \\
		\dhline
		CISCO Umbrella & Passive DNS measurements & \xmark & \cmark & $1M$ \\
		\dhline
		Cloudflare Radar & Passive DNS measurements & \xmark & \cmark & $\ge 1M$\\
		\dhline
		Majestic Million & Active Web crawls & Web & \cmark & $1M$\\
		\dhline
		Chrome UX Report & Browser telemetry & Web & \cmark & $1M$ \\
		\dhline
		HSTS Preload & Zone owner submissions & Web & \xmark & $126k$\\
		\bottomrule
 	\end{tabularx}
\end{table}

We refer to a set of curated or purposefully collected domain names that fulfill specific criteria as a \emph{hit list}.
Related work has already evaluated different aspects, such as reliability, reproducibil`ity, and fluctuation, of such lists and the implications of their use in research~\cite{Scheitle2018,LePochat2019,Ruth2022}.
Here, we only consider actively maintained and publicly available lists.
\autoref{tab:hitlists} provides an overview.

\paragraph{Tranco List}
An aggregation of various popularity-based top lists with the aim of being consistent, manipulation-resistant, reliable, and reproducible~\cite{LePochat2019}.
By default, domain names in the Tranco List are trimmed at the effective second level domain name (eSLD) and contain only the public suffix~(effective TLD; eTLD~\cite{eTLD2023}) plus the next label (referred to as \emph{pay-level domains} by list maintainers). 
For example \texttt{city.ofunato.iwate.jp} (an actual entry) consists of the public suffix \texttt{ofunato.iwate.jp} and \texttt{city} as its effective SLD.
This default structure can be modified through Tranco website\footnote{\url{https://tranco-list.eu}} where registered users can apply additional filters such as setting the combination algorithm or removing domains marked as unsafe by Google Safe Browsing database.

It should be noted that not all entries in this list resolve to IP addresses: at the time of this writing\footnote{List ID \texttt{Y5JYG}}, for example, 74 listed domain names out of the top 500 and 83,519 ($8.3\%$) of the top 1M do not resolve to an IP address (\ie have no DNS \texttt{A} records).
The main reason for this is the trimming behavior mentioned above that causes the list to include domain names belonging to CDNs, cloud service providers, or infrastructure operators, \eg \texttt{akamaiedge.net}, which do not resolve to IP addresses at that level.
Creating a custom list with subdomains beyond the eSLD partly solves this issue but results in inclusion of public suffixes such as \texttt{gob.es} or \texttt{ca.us}.
Beside a few exceptions~(\eg \texttt{gov.uk}), public suffixes do not resolve to IP addresses and some browsers~(\eg Chromium based) even consider them as a search query (and not a URL) when entered in the address bar.

Finally, the advertised options of filtering domains that resolve to an IP address, have a specific HTTP status, or return a minimum amount of HTTP content~(see original publication~\cite{LePochat2019}) were not available at the time of writing.

\paragraph{CISCO Umbrella}
This is a popularity ranked list based on passive measurements of DNS queries over CISCO Umbrella network\footnote{\url{https://umbrella.cisco.com}} (formerly OpenDNS services).
Similar to Tranco, reproducibility is provided (by date), but in contrast to the Tranco list entries are not trimmed to their functional SLD.

This list suffers from three handicaps.
First, about every fifth entry in the top 500 list (total of 118) does not resolve to an IP address.
Second, multiple subdomains of the same company or domain names sharing the same eSLD can cause measurement data skew.
For example, more than half of the top 25 entries (14 in total) are related to \texttt{netflix.com} or \texttt{netflix.org}.
Finally, opaque ranking and normalization algorithms used to generate this list are also among the drawbacks of this list.

\paragraph{Cloudflare Radar}
Similar to Cisco Umbrella, Radar statistics by Cloudflare are based on observations at own DNS resolvers (\texttt{1.1.1.1}).
The data can be fetched over Cloudflare website or through its API which provides more detailed information.
Beside a global ranking (over $1M$ domains), top 100 domains per country are also available.
Domain names in this set are trimmed at effective second level domain.

The downside of this list is twofold.
First, much like Tranco, trimming names at eSLD leaves a non-negligble portion of entries without a respective IP address.
And second, similar to CISCO umbrella, we observe high ranking names that are never directly visited by users, \eg \texttt{doubleclick.net} (advertising company).

\paragraph{Majestic}
The Majestic Million~\cite{Majestic2015} top list is a `link-level backlink index' (popularity ranked) list based on data gathered from web crawls.
The domain rank here (trimmed at eSLD) relates to \one the number of its `External Inbound Links', and \two counts of referring domain, IP, and subnets which are calculated during the crawl~\cite{Majestic2021}.

Among the top 500 domains in this list only 16 entries do not resolve to an IP address.
Invalid entries are either public suffixes (\eg \texttt{go.id}), private suffixes (\eg \texttt{azurewebsites.net}), or simply miss a label~(\eg \texttt{miit.gov.cn} without \texttt{www} label).

The main drawback of this list is its limited scope to the Web.
Using this as a point of departure to measure TLS in other contexts, such as SMTP, would require further steps (\eg query \texttt{MX} records) and would defeat the ranking: \texttt{facebook.com} (rank 1) is only popular as a website and not necessarily as a mail server.

\paragraph{Chrome UX Report}
As part of its user experience report, Google provides a popularity ranking of domain names based on data collected from Chrome browser users.
The ranking is further divided by country, platform, and popularity metric~\cite{Ruth2022a}.
In contrast to Tranco, host names are not trimmed at eSLD and Ranking is only provided `on a log10 scale with half steps'\footnote{See \url{https://developer.chrome.com/docs/crux/methodology/metrics}}.

The main advantage of this list is having browser data as its main source which leads to a better accuracy compared to other top lists~\cite{Ruth2022}.
For example, only 10 out of the top 1000 names in this list do not resolve to an IP address at the time of this writing.

\paragraph{HSTS Preload List}
Curated by Google as part of the Chromium Project, the `HTTP Strict Transport Security~(HSTS~\cite{RFC6797}) Preload List' is a collection of domain names that supporting browsers contact only through HTTPS\footnote{\url{https://www.chromium.org/hsts/}}.
At the time of writing more than 120k names are listed including public and private suffixes (\eg \texttt{zip}, and \texttt{now.sh}), eSLD domains, and individual domain names including all subdomains~(\eg \texttt{www.aclu.org}).
In contrast to the aforementioned hit lists, the entries here are not ordered and do not fulfill any specific criteria except supporting TLS.
Moreover, not every domain name in this list is delegated (\texttt{NXDOMAIN}) or carry an \texttt{A} record.

\subsection{Certificate Transparency Logs}
To address the problem of `misissued certificates', Certificate Transparency (CT) Logs~\cite{RFC6962,Scheitle2018b} were introduced.
These are publicly available and auditable data structures that are modified in an append-only manner.
CAs commonly log issued X.509 certificates in multiple CT logs allowing identity (\eg domain name) owners or CA subscribers to detect misissued certificates for their identities by monitoring CT logs.
Similarly relying parties, \eg browsers, make use of CT logs.
For example, Chrome and Safari browsers, with a combined global market share of over 90\%~\cite{BrowserShare2022}, validate server certificates only if they are logged in multiple CT logs~\cite{CTPolicy2023,ChromeCT2021,AppleCT2021}.
It is, thus, safe to assume that any certificate exchanged during a TLS handshake in a browser can be found in at least one CT log.
Even certificates that are used for purposes other than server and client authentication (\eg code signing) can partly be found in CT logs.
The corollary to this observation is that CT-logged certificates, or more specifically their Subject Alternative Names (of type DNS or IP), can be used as point of departure for TLS measurements.

\begin{figure}
    \scriptsize
    \input{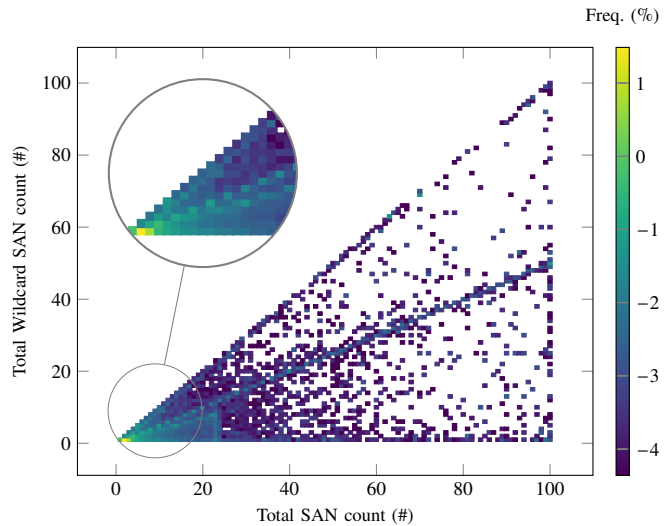}
    \caption{Frequency of valid certificates in our Tranco dataset based on the number of wildcard and total SAN of type DNS in each certificate (capped at 100)}
    \label{fig:san-wildcard-freq}
\end{figure}

Note that due to wildcard certificates, it is not possible to discover every viable domain name from CT logs.
In our \texttt{TRANCO} dataset (see \autoref{sec:appendix}), for example, about $40\%$ of all valid certificates have at least one wildcard SAN while less than $1\%$ solely include wildcard names.
\autoref{fig:san-wildcard-freq} depicts the frequency of certificates in our \texttt{TRANCO} dataset based on number of wildcard SAN entries and the total number of SANs in each certificate.

Finally, it should be noted that not every logged certificate is also deployed on a TLS server.
Cloudflare, for example, vicariously applies for two certificates for its customers: a production and a backup certificate to instantly replace it in emergency cases~(\eg key compromises)~\cite{CloudflareBackup2022}.
Furthermore, it is not possible to infer which service actively use a logged certificate~(\eg HTTPS or SMTP).
Finally, for a given logged certificate (even if not expired or revoked) it is not guaranteed that the domain names that it covers is still delegated or resolves to an IP address.

\subsection{DNS Zone Files}
\label{subsec:src-dnszone}
The Internet Corporation for Assigned Names and Numbers (ICANN) allows interested parties to directly request generic TLD (gTLD) registries for their zone files through the Centralized Zone Data Service (CZDS).
Zone files for TLDs are commonly used in DNS and DNSSEC measurements~\cite{Zhu2015,Osterweil2022,Lee2022} but are also suitable in context of TLS and Web PKI (see, for example, Holz~\etal~\cite{Holz2019}).

The downside of using DNS zone files is twofold: \one not every TLD is part of CZDS and not every request is granted, and \two the massive amount of names \textemdash\texttt{.com} alone is estimated to have 160M delegations~\cite{TLDCount2023}\textemdash in no specific order poses a challenge for both the measurement procedure and interpretation.
Merits of using DNS zone file as point of departure have been discussed in related work~\cite{Zhu2015}.

\subsection{Indexing Services}
Numerous commercial companies in the field of Internet measurement and scanning provide access to their data which can be used as a point of departure for TLS and Web PKI measurements.
Two prominent examples are Censys Search\footnote{\url{https://censys.io/data-and-search/}} and Rapid7 Open Data\footnote{\url{https://opendata.rapid7.com}}:
Censys performs daily scans of the IPv4 address space and provides an API to query and search collected data.
Data by Rapid7 is limited to its Open Data repository containing artifacts of regular scans by Project Sonar~\cite{Rapid7} and lacks search and query features.

Downside of using such services as data sources is lack of control and transparency on underlying scanning procedures.
Wan~\etal~\cite{Wan2020}, for example, discuss how regular scans by Cencys (using ZMap; see \autoref{sec:tooling}) leads to its source IP addresses being blocked and consequently missing up to $4.6\%$ of all HTTPs hosts.
A similar phenomenon is also observed by Chung~\etal~\cite{Chung2016} for datasets from Rapid7 and University of Michigan, suggesting that data from regular scanners are susceptible to being incomplete due to blacklisting.

\subsection{Internet Traffic Monitors}
Instead of collecting data, we could rely on data collected by traffic monitors.
Passive measurements present an authentic snapshot of Internet usage but are usually limited in their scope, \eg traffic on a university network, and require access to special network nodes.

Note that although using passive measurements for TLS and Web PKI analysis is widespread~\cite{Holz2011,Amann2017,Gustafsson2017,Kotzias2018,Ouvrier2017,Holz2016}, such data might not be suitable for specific use cases~\cite{Zhu2015,Zhang2021}.

	\section{Tooling}
\label{sec:tooling}
\begin{table*}
	\caption{Overview of selected TLS measurement tools and their capabilities}
	\label{tab:tools}
	\begin{tabularx}{\textwidth}{X >{\ttfamily}c c c c c c c c l c c c c c}
		\toprule
		& & \multicolumn{2}{c}{Command Line Interface} & & \multicolumn{4}{c}{TLS$^\dagger$} & & \multicolumn{5}{c}{X.509$^\ddagger$} \\
		\cmidrule{3-4} \cmidrule{6-9} \cmidrule{11-15}
		Tool & \textnormal{API} & Batch Process & Machine Readable & & T1 & T2 & T3 & T4 & & X1 & X2 & X3 & X4 & X5 \\
		\midrule
		OpenSSL & C & \xmark & \xmark & & \Circle & \CIRCLE & \CIRCLE & \Circle & & \Circle & \LEFTcircle & \CIRCLE & \CIRCLE & \LEFTcircle \\
		\dhline
		curl & C & \xmark & (JSON) & & \Circle & \LEFTcircle & \LEFTcircle & \CIRCLE & & \Circle & \LEFTcircle & \LEFTcircle & \LEFTcircle & \Circle \\
		\dhline
		ZMap Suite & GO & \cmark & JSON & & \Circle & \CIRCLE & \LEFTcircle & \CIRCLE & & \CIRCLE & \CIRCLE & \CIRCLE & \LEFTcircle & \Circle  \\
		\dhline
		TLS-Attacker Suite & Java & \cmark & DB & & \LEFTcircle & \CIRCLE & \CIRCLE & \Circle & & \Circle & \Circle & \CIRCLE & \Circle & \Circle \\
		\dhline
		testssl.sh & \xmark & \cmark & CSV / JSON & & \LEFTcircle & \LEFTcircle & \LEFTcircle & \LEFTcircle & & \Circle & \LEFTcircle & \Circle & \CIRCLE & \LEFTcircle \\
		\dhline
		goscanner & \xmark & \cmark & CSV & & \Circle & \CIRCLE & \Circle & \CIRCLE & & \Circle & \CIRCLE & \Circle & \Circle & \Circle \\
		\dhline
		Puppeteer & NodeJS & \multicolumn{2}{c}{--- N/A ---} & & \Circle & \Circle & \Circle & \CIRCLE & & \Circle & \CIRCLE & \LEFTcircle & \Circle & \Circle  \\
		\bottomrule
	\end{tabularx}
	\begin{minipage}{\textwidth}
		\vspace{2pt}
		\scriptsize
		$\dagger$ \textbf{T1}: verifies conformity to TLS standards / \textbf{T2}: can manipulate handshake layer messages / \textbf{T3}: can manipulate record layer messages / \textbf{T4}: implements relevant application layer protocols\\
		$\ddagger$ \textbf{X1}: verifies conformity to to X.509 standards / \textbf{X2}: can validate certificate similar to a browser (except proprietary CRL lists) / \textbf{X3}: can fully parse certificate / \textbf{X4}: can evaluate revocation status / \textbf{X5}: can relate the certificate to DNS records\\
		\CIRCLE~Full support / \LEFTcircle~Partial support / \Circle~No support
	\end{minipage}
\end{table*}

To draw a holistic picture of TLS and Web PKI numerous aspects need to be measured.
Nearly all general-purpose programming languages offer implementations of common Internet protols, \eg TLS and HTTP.
However, not all are appropriate for measurement purposes.
For example, the standard crypto library of Node.js, a JavaScript engine, is limited to parsing selected fields from an X.509 certificate, and implementing client-side session tickets is a tedious task over its TLS API.

In this section we introduce some widely established tools for measuring TLS and Web PKI and discuss their merits regarding features introduced in \autoref{sec:meas}.
\autoref{tab:tools} provides an overview.

\paragraph{OpenSSL}
An open source implementation of a wide variety of cryptographic algorithms and protocols, OpenSSL is used in client (\eg curl) and server software (\eg NGINX) alike and lends itself as an appropriate tool to collect and analyze TLS and Web PKI data.

It can be used programmatically through its C API or any of its CLI tools, such as \texttt{s\_client} (a TLS client), or \texttt{x509} (an X.509 certificate parser).
However, manipulating low level details, \eg of handshake (\ref{itm:T2}) and record layers (\ref{itm:T3}) or X.509 extensions (\ref{itm:X3}), is limited to its API.
OpenSSL is capable of validating signed certificate timestamps (CT) (related to \ref{itm:X2}) and DANE records as well as revocation status over CRL, OCSP, and stapled OCSP (\ref{itm:X4}) if required resources, \eg DNS resource records, are provided additionally.

OpenSSL suffers from four shortcomings: \one low level API (\eg manual memory management), \two non-machine-readable CLI output, \three partial nonconformity with RFCs (\ref{itm:T1}), \eg OSCP response is not validated if signed by a designated entity and a complete certificate chain is missing (see RFC2560 guidelines~\cite[\S 4.2.2.2]{RFC2560}).
Standard distributions of OpenSSL also do not contain weak cipher suites which are relevant for \ref{itm:T2} and \ref{itm:T3}.
And \four higher level protocols are not implemented (\ref{itm:T4}).

\paragraph{curl}
A command-line tool and library (\texttt{libcurl}) that implement various Internet protocols.
curl is extensive in terms of configuration and protocol support (\ref{itm:T4}) but has its focus mainly on transferring data.
As such, the CLI only provides a rudimentary option to print data in machine-readable format: either by selecting from a set of predefined keys or by printing all the keys in JSON format.

The drawbacks are as follows: \one specific data, \eg raw certificates, as well as granular access to handshake and record layers are (\ref{itm:T2} and \ref{itm:T3}) not given over the CLI but only through \texttt{libcurl}.
This API, however, is not as extensive as the underlying TLS engine, \ie provides limited callbacks to OpenSSL functions or available cipher suites.
And, \two curl is limited in parsing X.509 certificates and provides unstructured data over the CLI and uses its own structure over the API (\ref{itm:X3}), \three it does not support validation of SCT and OCSP (\ref{itm:X4}), and finally \four it cannot be configured to validate DANE (\ref{itm:X5}).
By default, curl distributions are based on OpenSSL and inherit all respective shortcomings as discussed above.
It is, however, possible to manually build curl with other TLS engines.

\paragraph{ZMap Project}
ZMap is a network scanner similar to \texttt{nmap} but with a focus on modularity and speed~\cite{Durumeric2013}.
It can be used both programmatically and over the CLI.
Since its introduction, ZMap has been grown into a software suite including, among others, an application-layer scanner (ZGrab), a DNS lookup tool (ZDNS), and an X.509 certificate parser (ZCertificate) and linter (ZLint).

ZMap CLI tools produce machine-readable output in JSON format and can be combined using standard Linux pipes.
For example combining \texttt{zmap}, \texttt{zls}~\cite{Izhikevich2021}, and \texttt{zgrab2} allows scanning hosts for various ports, detect services, and finally perform an application layer handshake.

ZMap accepts custom client hello handshakes and allows modifying record layer through cipher suites selection (\ref{itm:T2} and \ref{itm:T3}).
It also supports various application layer protocols using (opportunistic) TLS (\ref{itm:T4}).
X.509 certificates can be checked against standards, such as RFCs and ETSI, as well as other relevant policies, such as CA/B baseline requirements and browser PKI/CT policies (\ref{itm:X1}).
It also implements certificate validation similar to common browsers but without integrating proprietary CRL lists (\ref{itm:X2}).

The shortcoming of ZMap Suite is twofold: \one it only supports stapled OCSP and not OCSP or CRL (\ref{itm:X4}), and \two it lacks built-in means to correlate DNS and X.509  to relevant DNS records (\ref{itm:X5}).

\paragraph{TLS-Attacker Suite}
A collection of tools and libraries in Java with a focus on security analysis of TLS servers and clients.
Similar to ZMap it is capable of performing Internet-wide TLS scanning and vulnerability detection (\ref{itm:T1}).
It also allows for fine-grained manipulation of TLS handshakes and exchanged messages (\ref{itm:T2} and \ref{itm:T3}), and provides tools to parse and generate X.509 certificates (\ref{itm:X3}).
Support for Web PKI aspects, however, are rather limited (see \autoref{tab:tools}).

\paragraph{testssl.sh}
A portable script to analyze security of SSL/TLS servers, testssl.sh checks for known TLS vulnerabilities (\ref{itm:T2} and \ref{itm:T3}), and simulates TLS handshakes for a variety of known clients (\eg Firefox 100 on Win 10).
It also supports automatic certificate validity checking over CRL and OCSP endpoints (\ref{itm:X4}), and looks for DNS CAA records (without matching however).

This tool is rather suitable for administrators than performing measurement at large while lending itself as a good alternative for debugging single noticeable observations from measurement datasets.
Disadvantages of testssl.sh can be summarized as follows: \one its lack of an API, and its monolith design that does not support modification without changing the core codebase, \two insufficient certificate information (\ref{itm:X3}) and lack of an option to store the full certificate chain, and finally \three file storage for results causes slow-downs in high concurrency measurements due to file system operation overhead (see \autoref{sec:perform}).

\paragraph{Goscanner}
This is command line tool written in Go language.
It has the ability of fingerprinting TLS handshake using various methods~\cite{Althouse2020,Sosnowski2023} (\ref{itm:T2}), validating and storing X.509 certificate chains (\ref{itm:X2}), and using custom handshakes when contacting servers.
The output is stored in multiple files in CSV format.

\paragraph{Puppeteer}
A Node.js library (JavaScript) that enables programmatic control of browsers which support Chrome DevTools Protocol (CDP)\footnote{\url{https://chromedevtools.github.io/devtools-protocol/}}.
Although limited to the Web, measurements using Puppeteer can precisely simulate user experience (see \autoref{sec:interpret}).
It can, for example, properly follow redirects of various types (\eg HTML and JavaScript) and extract certificates as presented by the browser to users.

Puppeteer, however, is limited in scope by four factors: \one the CDP protocol does not provide access to lower level API and many features are still experimental (\eg retrieving certificates), \two Chromium configuration flags that define its behavior upon initiation are not well-documented, \three page navigation (equivalent to opening a tab in browser) spawns multiple processes and consumes a relatively high amount of resources (computation and memory).
As the number of concurrent Puppeteer jobs increases, Puppeteer fails to properly kill Chromium process and free up memory.
Finally, \four underlying browsers used by Puppeteer might exhibit non-standard behavior.
Chrome, for example, artificially generates an HTTP redirect (status code \texttt{307}) when encountering an HSTS header even if the server does not explicitly indicate a redirect.

\paragraph{Zeek}
Formerly known as Bro, Zeek is an online monitoring and analysis tool that is commonly used for passive measurements.
It has its own scripting language that allows for customizing its behavior.

	\section{Preparing Measurements}
\label{sec:prepare}
Depending on the measurements goals, we need to take different aspects into considerations before initiating the actual measurement.
Here we address getting familiar with best practices, and choosing appropriate vantage point and measurement features, and subsequently provide a brief discussion on proper documentation for reproducibility, and software/hardware validation in advance.

\subsection{Best practices}
Performing measurements, specifically large-scale or Internet-wide, can be a source of disturbance, \eg can be regarded as malicious attacks to infrastructure operators.
Here we discuss two aspects that must be considered before performing a measurement.

\paragraph{Ethical considerations}
Durumeric, Wustrow, and Halderman~\cite[\S 5]{Durumeric2013} recommend 7 practices that address ethical aspects of Internet-wide measurements ranging from coordinating with local network admins to information provision at source DNS and IP addresses, and catering for simple opt-outs.
Partridge and Allman\cite{Partridge2016} go beyond technical means and discuss `tangible harm to people' that can be caused by active measurements.
In a recent paper, Pauley and McDaniel\cite{Pauley2023} provide an overview of previous work of ethical measurement, summarize existing community guidelines, and discuss recommendations to establish an `ethical framework' for Internet measurements.

\paragraph{Responsible disclosure}
If security vulnerabilities are detected during measurements, researchers are encouraged to inform affected entities through \emph{Coordinated Vulnerability Disclosure} (aka Responsible Disclosure).
Governmental organization (\eg US CISA or Germany's BSI), standardization institutes (\eg NIST or ETSI), and other relevant entities have their own guidelines and procedures to document and submit vulnerabilities.
The \emph{CERT Guide to Coordinated Vulnerability Disclosure}\cite{Householder2019} is, for example, a comprehensive guide by Carnegie Mellon University.

\subsection{Vantage Points}
The physical location where measurements are performed can have a twofold impact on the results in terms of reachability, and consistency.
As such, same measurement (method and target) from different vantage points can lead to different resuls.

\paragraph{Reachability}
Wan~\etal~\cite{Wan2020} discuss how host reachability can vary depending on the vantage point.
They show how the origin location in single-probe IPv4 address space scans can lead to missing around 5\% of all HTTPS hosts in the worst case (See Censys in \autoref{sec:datasrc}).
Reachability itself, however, can be subject of measurement instead of being a mere measurement artifact, \eg in censorship studies~(see, for instance, Raman~\etal~\cite{Raman2020a,Raman2020}).
In such cases it is desired to measure data from specific vantage points to study reachability, even by crowd-sourcing the measurements through platforms such as OONI~\cite{Filasto2012} or RIPE Atlas~\cite{Staff2015} to avoid detection or blocking.

\paragraph{Consistency}
Depending on the vantage point, the same domain name can be resolved and routed to different IP addresses, \eg to the nearest point of presence of a cloud provider.
Lee~\etal~\cite{Lee2021} show how different edge servers hosting content for the same domain name can be configured differently with respect to TLS security.

\subsection{Measurement Features}
We need to define the feature sets that we want to record for each host when performing a measurement.
For example, if we are only interested in X.509 certificates, TLS record messages and any application layer data can be ignored while further information such as DANE records or CRL and OCSP data needs to be fetched in addition.

These features are related to entities depicted in \autoref{fig:overview} and can roughly be categorized as technical and non-technical.
The former category  is summarized in \autoref{sec:meas} and comprises \one TLS handshake and record layer features, \eg supported cipher suites, \two DNS related resource records, \eg TLSA records and DNSSEC chain, and \three Web PKI entries, \eg X.509 chains and OCSP signer certificates.
We categorize other features that are not manifested in technical terms but might be relevant for analysis, interpretation of data, or reproducibility of results as non-technical.
For example, applicable CP/CPS documents of CAs or government regulations at the time of study belong to this category (see \autoref{sec:interpret}).

\subsection{Documentation for Reproducibility}
To establish measurement results, it should be reproducible, \ie it should be possible to reach to the same results using the same procedure or artifacts~\cite{ACM2020}.
We refer to related work~\cite{Bajpai2017,Bajpai2019,Demir2022,Cangialosi2022} for an extensive discussion of best practices and challenges of reproducibility.
Here, we confine ourselves to a brief overview regarding software and data.

\paragraph{Software}
When using off-the-shelf software (see \autoref{sec:tooling}), the exact version must be noted.
For open source software a reference to specific state, \eg a git branch or commit, is preferable.
Any modification to the software must also be documented, \eg using inline comments, accompanied by a short reason, \eg bug in original software.

Hard coding configurations should be avoided, and measurement parameters should be externalized in dedicated files.
Similarly, the list of dependencies (software and system alike) should be documented and provided in a readily installable format, \eg \texttt{pip freeze} and \texttt{pip install -r} for Python software.
Optimally, a configure script (\eg generated by GNU Autoconf) can help other researchers to check their systems for missing libraries and other dependencies.

Instruction should be provided on how to start, configure, and optimize each program for measurement.
Respectively, structure and content of generated output and error codes must be disclosed.
A step-by-step manual to execute a complete measurement must also be available.
To reproduce a specific measurement exact configuration and input parameters (\eg list of measured hosts) alongside deployment settings (\eg vantage point) are to be provided.

Source code, configurations, and even dependencies can be bundled together, \eg in a container image, to cater for easier and faster bootstrapping.
Some tools also allow defining virtual environments that can be exported and imported on any arbitrary system.

\paragraph{Data}
A measurement might consist of multiple phases where output from one phase is fed as input to another phase.
Moreover, utilized software might implicitly (\eg due to convention over configuration) rely on files that differ from one execution environment to the other.
For example, OpenSSL is shipped with a set of default configuration\footnote{Can be printed using \texttt{openssl version -d} command} (\eg path to trusted CAs) that impact certificate validation and other operations.
Both implicit and explicit data (files) are required for reproducibility.
Each piece of information should be accompanied by metadata describing its content, structure, timestamps, etc.

The granularity of data is also relevant.
Instead of providing only end results, having interim data for each step of the measurement allows others to verify correctness of the overall procedure.
For example, in place of stating if a host supports stapled OCSP, TLS handshake containing OCSP response alongside the CA certificates (from local trust store) used to validate it should be provided.

\subsection{Validation of Software and Hardware}
The final step in preparing measurements is to validate the software and hardware before executing the complete measurement.
Wan~\etal~\cite{Wan2020}, for example, runs their measurement in small scale (1\% or IPv4 address space) from all selected vantage points to confirm that both software and hardware can reach desired scanning speed while avoiding extraordinary packet drops.

	\section{Performing Measurements}
\label{sec:perform}

After we have determined our source and a set of features, the actual measurement can be started.
In this section, we discuss the importance of maintaining temporal integrity while collecting data from different sources, different types of stateful measurements, detecting and handling errors during measurement, and finally how to best store data.

\subsection{Temporal Integrity}
\begin{table}
	\scriptsize
	\caption{Selected measurement features which are temporally bound and are collected at various sources}
	\label{tab:validity-features}
	\begin{tabularx}{\columnwidth}{X l l l}
		\toprule
		Type & Source & Collected at & Temporal indicator \\
		\midrule
		DNS RR & Auth. NS & DNS resolution & TTL field$^\dagger$ \\
		\dhline
		DNSSEC \texttt{RRSIG} & Auth. NS & DNS resolution & Inception / expiration fields$^\ddagger$ \\
		\dhline
		X.509 cert & Server & TLS handshake & Not before / not after fields$^\ddagger$ \\
		\dhline
		TLS session & Server & TLS handshake & Lifetime hint$^\dagger$ \\
		\dhline
		Stapled OCSP & Server & TLS handshake & This update / next update$^\ddagger$ \\
		\dhline
		OCSP & CA & CA OCSP endpoint & This update / next update$^\ddagger$ \\
		\bottomrule
		\multicolumn{4}{l}{$^\dagger$ Relative / $^\ddagger$ Absolute}
	\end{tabularx}
\end{table}

There are numerous pieces of temporally bound data that we might collect during our measurement from various sources: \autoref{tab:validity-features} provides an overview.
To make sure that our measurements capture a correct snapshot, we must make sure that we measure these features with short temporal discrepancies.
That is to reduce the time distance between probing related features, for example between querying DNS \texttt{A} records and establishing the TLS connection, or fetching certificates and validating OCSP response.
Nawrocki~\etal~\cite{Nawrocki2022}, for example, notes how a short time difference between two measuring X.509 certificates can result in collecting different certificates for the same host due to key roll-overs.
Zirngibl~\etal~\cite{Zirngibl2021} observe in their analysis of QUIC how Google rotates certificates for \texttt{googlevideo.com} \emph{even} before these are expired.
Finally, in their study of DANE use by SMTP servers, Lee~\etal~\cite{Lee2022} notes the necessity of maintaining temporal integrity and collects DANE records and X.509 in an atomic operation.

\begin{figure}[b]
	\scriptsize
	\centering
	\input{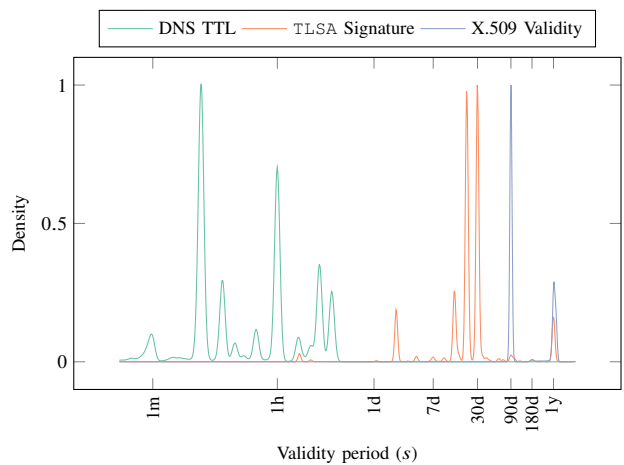}
	\caption{Validity period of X.509 certificates, DNSSEC signature validity of \texttt{TLSA} records, and actual remaining TTL of \texttt{A} records at the time measurement from our \texttt{TRANCO} dataset}
	\label{fig:validity-periods}
\end{figure}

\autoref{fig:validity-periods} depicts temporal values for different records that we collected for our \texttt{TRANCO} dataset.

\subsection{Stateful Measurement}
Some aspects of TLS cannot be analyzed based on one-off and stateless measurements.
In the following we discuss cases that require measuring same servers multiple times or measuring different servers in the same context and with respect to each other.

\paragraph{Same host and settings}
Here, for each host in the sample set multiple measurements are performed.
For example, Liu~\etal~\cite{Liu2015} observe that some servers deliver stapled OCSP only if it has already been cached locally~\cite[\S 4.3]{Liu2015}, so to determine if Stapled OCSP is supported, multiple successive measurements are to be performed.
In a later study~\cite{Chung2018}, the authors expand their scope to investigation of OCSP responders using hourly measurements in a longitudinal study (4 months period) from multiple vantage points, practically forcing servers to always query and cache OCSP responses.

\paragraph{Same host with different settings}
In other cases, we might need to alternate measurement settings and parameters.
For example, discovering how TLS sessions are implemented and deployed requires performing multiple handshakes with and without session IDs or tickets.
In a study from 2018, Sy~\etal~\cite{Sy2018} regularly establish TLS connections to Alexa Top 1M sites and show how TLS resumption can be used to track visitors.
Hebrok~\etal~\cite{Hebrok2023} examine cryptographic issues of TLS resumptions by executing 10 TLS handshakes for every (virtual) host while storing session keys alongside session tickets.
Furthermore, collected session tickets were modified to check for server-side authentication vulnerabilities.

\paragraph{Different hosts with same settings}
In some cases, we need to perform subsequent measurements based on information retrieved during an initial measurement of a sample.
For example to study TLS sessions that are shared among virtual hosts at the same or among different servers.
Sy~\etal~\cite{Sy2020} show performance advantages of sharing session among domain names given that they are all controlled by the same entity. 
Similarly, Springall, Durumeric, and Halderman~\cite{Springall2016} study shared TLS session among different domain names but set the focus on security ramifications instead of performance.
For this, session tickets were collected regularly over a period of 9 weeks and finally compare with each other to detect, among others, security risks of servers reusing session cache, session ticket, and Diffie-Hellmann parameters.

\paragraph{Different hosts with different settings}
Finally, there are cases that an initial measurement of a single sample are used to define and perform measurements on related samples but under different settings.
An example is the family of TLS confusion attacks.
This is an umbrella term referring to a type of MitM attacks where an attacker manages to divert TLS traffic from intended server to another without the client noticing.

Delignat-Lavaud and Bhargavan~\cite{DelignatLavaud2015} show how a TLS terminating proxy server can be tricked into redirecting HTTPS traffic meant for a virtual server to another.
The anecdotal examples of this study are extended by the systematic analysis of Zhang~\etal~\cite{Zhang2020} where for a set of host names sharing an X.509 certificate TLS connections are made to different IP addresses and ports.
Brinkmann~\etal~\cite{Brinkmann2021} introduces a cross protocol variation of TLS confusion attacks where traffic is diverted from one protocol to another, \eg HTTPS to SMTPS.

\subsection{Failure handling}
Misconfigurations, blackouts, deficient workflow, and software bugs are among causes for measurement errors and failures.
It is, thus, integral to detect errors early on and subsequently be capable of resuming measurements without having to start atop.

\paragraph{Detecting sources of error}
Nearly all tools discussed in \autoref{sec:datasrc} support debug mode which can be used to print extra operational information to standard error stream.
Debug and error messages, however, are not always in machine-readable format or include necessary information (\eg passed arguments) to detect causes for failure.
Another (complimentary) approach is to dump data packets in transit during measurements, \eg using Wireshark, and to dissect them in case of failures that cannot be debugged otherwise.

\paragraph{Detecting measurement artifacts}
In addition to technical failures, such as software bugs, that manifest themselves in error logs, software crashes, \etc, deficiencies in measurement techniques can lead to measurement artifacts.
This kind of error is harder to detect and commonly materialize itself after inspecting measurement results.
For example consider the measurement agent being marked as bot by cloud providers or having a TLS connection terminated due to lack of SNI.
In both cases the logs will not indicate any failures, but the software has practically failed to capture a truthful snapshot.
Although not all measurement artifacts can be avoided, it is important to monitor measurement for any cases that might serve as an indication.

\paragraph{Resuming measurements}
Specially in case of large-scale measurements or limited resources, it is desirable to be able to resume failed measurements, \ie skip successful measurements, rerun failed ones, and continue with the rest.
None of the tools with built-in batch process (see \autoref{tab:tools}) supports measurement resumption.
To address this shortcoming, tools can be wrapped inside process managers, \eg GNU Parallel~\cite{Tange2011}, that are able to parallelize and monitor processes and cater for resumption in cases of failures.

\subsection{Data Storage}
Measurement data can are commonly stored in traditional databases (\eg TLS-Attacker), flat-file databases (\eg Z-Map suite), or simple files (\eg per certificate, per TLS handshake, \etc).
Storing each feature in a single file simplifies searching and filtering but the added overhead due to file system and disk I/O operations can lead to throttling measurement speed and parallelization.

In practice, a mixture of all strategies might be chosen.
Holz~\etal~\cite{Holz2016}, for example uses Z-Map to discover viable hosts, collects data using OpenSSL and stores the result in a database.

	\section{Interpreting Data}
\label{sec:interpret}
After designing and performing measurements, we need to interpret data.
In this section we discuss a number of pitfalls that can skew measurement-based conclusions and insights.

\subsection{Certificate Subjects}
It might be useful to detect the subject of a certificate, for example, to detect if a certificate is being shared among distinct service providers~\cite{Liang2014,Cangialosi2016}.
Depending on validation type (DV, OV, IV, and EV; see \autoref{sec:bg}) corroborating data might be needed to uniquely identifying a certificate owner.

For EV certificate, the subject name explicitly denotes the entity running a website, \ie the organization responsible for its content~\cite[\S 2.1.]{CABEV} and is sufficient to extract the certificate owner.
In contrast, OV and IV certificates identify the certificate applicant~\cite[\S 3.2.2.]{CABReq2.0}.
In case of \emph{delegated services}~\cite{Liang2014}, \eg when a CDN operator vicariously applies for a certificate, the subject might actually identify the delegatee and not the delegator and cause a misinterpretation of the certificate.
Cloudflare, for example, uses OV certificates with its own identity information for its free-tier customers ---a marketing stunt that can be avoided by upgrading to paid plans~\cite{Cloudflare2023}.

DV certificates pose the biggest challenge, as the only subject information is provided in form of SANs (domain name, IP addresses, etc.) and cannot be used to identify the certificate owner.
Although SAN entries are supposed to be \emph{alternative names} of the same subject, there is no guarantee that all entries actually denote the same owner.
Multitenant infrastructure operators, \eg Google and Imperva,  list domain names of different customers as SANs of the same certificate.
To address such issues, Cangialosi~\etal~\cite[\S 4.1]{Cangialosi2016}, for example, devise a method based on e-mail addresses included in DNS WHOIS data to infer if two domain names belong to the same organization.
This method, however, might be less fruitful for future measurements as information in WHOIS databases are more restricted nowadays due to privacy concerns and regulations such as EU GDPR.

\subsection{Certificate Issuers}
Each certificate denotes its issuing and signing CA in its issuer field.
Understanding the role of CAs in the Web PKI has also been part of research.
Durumeric~\etal~\cite{Durumeric2013a}, for example, give a detailed overview of CAs (\eg country of origin and market share) alongside an analysis of which type of organizations (\eg libraries, museums) are awarded with unconstrained intermediate CA certificates (see \autoref{sec:bg}).
Fadai~\etal~\cite{Fadai2015} go further and investigate CAs in terms of  corruption, human rights, \etc at their respective country of origin.

\begin{figure}
    \includegraphics[width=\columnwidth,keepaspectratio]{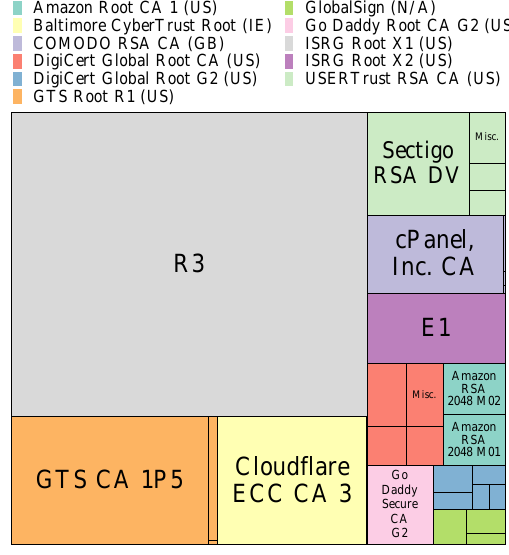}
    \caption{CA market share from our Tranco dataset. Each colored box denotes a trust anchor (root certificate) and nested boxes are intermediates. \emph{Misc.} denotes an aggregated share of multiple intermediates. The size of each box is directly proportional to the number of leaf certificates issued by respective intermediate CA. Some labels are removed due to space constraints.}
    \label{fig:ca-marketshare}
\end{figure}    

Identifying the operator of a CA using only its (subject) name is, however, a non-trivial task as Ma~\etal~\cite{Ma2021a} notes.
The authors show, for example, how a given root CA does not refer to its operator, DigiCert, in its subject name (\texttt{CN=Hotspot 2.0 Trust Root CA - 03; O=WFA Hotspot 2.0;C=US}).
Looking at DigiCert PKI repository~\footnote{\url{https://www.digicert.com/kb/digicert-root-certificates.htm}}, also reveals a number of other root CAs operated by DigiCert but under different subject names such as Baltimore CyberTrust Root, GeoTrust Primary Certification Authority, \etc
The respective impact on research data interpretation is twofold: \one analysis that based on identity of CAs requires additional care.
For example, 5 out of the 14 listed CAs in the market share statistics by Aas~\etal~\cite[\S 7]{Aas2019} all belong to DigiCert.
And \two knowing the actual operator behind CAs helps understanding specific trends.
For example, why all banks in our \texttt{IR-BANKS} set are subscribers to an obscure CAs (Unizeto) as we discuss in the next subsection.

\autoref{fig:ca-marketshare} depicts the market share of CAs from our \texttt{TRANCO} dataset divided by trust anchors (colored boxed) and intermediate CAs (nested boxes).
The size of each box is directly proportional to the number of leaf certificates issued by respective intermediate CA.
More than half of certificates are issued by Let's Encrypt intermediates R3 and E1, respectively under trust anchors ISRG Root X1 and ISRG Root X2.
There are also trust anchors that are exclusively used by a single company, \ie GTS Root (Google), Amazon Root CA, and GoDaddy Root CA.
We also observe PKI operators such as USERTrust and DigiCert that provide PKI as-a-service and manage intermediate CAs for third parties.
It is also notable that all root certificates in this set are under US jurisdiction, even Baltimore CyberTrust Root which denotes Ireland as country in its subject field is part of DigiCert since 2015.

\subsection{CA Operational Policies}
All relevant information from applicability of a certificate to its issuance, revocation, and definition of custom OIDs are defined in CP and CPS~\cite{ITUX.509,RFC3647} of its CA.

These documents might be helpful in better understanding abnormalities in measurements.
For example, all Iranian banks (in \texttt{IRB} set) use certificates issued by \textit{Unizeto Technologies}, a Polish CA, while none of the banks in our \texttt{ECB} set (except the \textit{PKO Bank Polski} in Poland) subscribe to that CA.
Consulting CP and CPS document for the QuoVadis and DigiCert (the former being a subsidiary of the latter) that cover more than half of all European bank and credit institutes reveals that DigiCert is prohibited by the US law to provide any services to countries or companies on a \emph{government denied list}~\cite[\S 4.1.1]{QuoVadis2023}, \eg Iran.

When considering CP and CPS documents, three aspects should be regarded.
First, although RFC3647~\cite{RFC3647} streamlines the structure of CP and CPS, the terminology and concepts applied might vary among CAs.
Second, in practice CAs do not necessarily behave in conformance with their own or other relevant guidelines.
For example, the \textit{ECC CA-3} intermediate CA certificate from Cloudflare\footnote{\url{https://crt.sh/?id=2392142533}} carries OIDs for DV, OV, and IV validation types, and violates both its issuer CP~\cite[\S 3.2.2]{DigiCert2023} (IV certificates are issued to individuals) and CA/B Baseline Requirements~\cite[\S 7.1.2.7.3]{CABReq2.0} (IV certificates must carry \texttt{surname} and \texttt{givenName} in their subject field).
And third, there is no guarantee of logical coherence.
Let's Encrypt, for example, asserts no relationship between subscribers and registrants of domains names in a certificate~\cite[\S 3.1.4]{ISRG2023} in its combined CP/CPS and in the same document requires subscribers to be the legitimate registrant of respective domain names~\cite[\S 9.6.3]{ISRG2023}.

\subsection{Certificate Validity}
Authentication presupposes presentation of a valid certificate.
Beside formal requirements (see RFC5280~\cite{RFC5280}), a relying party dictates the validation methods and validity criteria.
The most basic form of validating a certificate is \emph{path validation} described in RFC5280~\cite{RFC5280}: validation succeeds if certificate subject matches the desired subject and if there is trusted path to a trust anchor (root).
Depending on the relying party (user) or the client it uses (\eg a web browser), the set of trust anchors and validation policies might differ.
Respectively, to verify if a collected certificate is valid three features needs to be defined: \one trust store, \two revocation status, and finally \three additional policies.

\paragraph{Trust Stores}
Major browsers, operating system vendors, governments, and even individual organizations maintain sets of \emph{trustworthy} certificates (intermediate and root CAs; see \autoref{fig:overview}).
Although users are generally oblivious to their role, trust stores are integral in securing or jeopardizing the security~\cite
{Soghoian2010,WP2022} of users daily communications.

Each trust stores maintainer has its own set of requirements for incorporating a CA: for example, through consensus as it was the case for Debian OS (\texttt{ca-certificates} package~\cite{Shuler2011}), well-defined policies (including undergoing audits and conforming to requirements) as in Mozilla Root Store Policy~\cite{Mozilla2023}, or by establishing a legal framework as in the EU eIDAS Regulation~\cite[Article 24]{EUeIDAS2014}.
Related work has extensively studied root stores and their relation to each other~\cite{Perl2014,Fadai2015,Korzhitskii2020,Ma2021,Purushothaman2022}.

Beyond established trust stores, users may also decide to trust CAs by adding them manually or through software installations.
It is, for example, common practice for online banking software in South Korea to install additional CAs in local trust stores~\cite{Palant2023}.
The impact of such settings can also be observed in TLS measurements.
A study from 2016~\cite{Chung2016}, for example, notices Korea Telecom as one of top Autonomous Systems that host invalid certs presumabely because the researchers rely on generic trust store for validations.
Zhang~\etal~\cite{Zhang2021} study the prevalance of such `hidden root CAs' within the Web PKI ecosystem.

\paragraph{Revocation Status}
A certificate that fulfills path validation might still be invalid if it has been revoked.
Depending on the issuing CA, relying parties can check the revocation status using Certificate Revocation Lists (CRL)~\cite{RFC5280} or the Online Certificate Status Protocol (OCSP)~\cite{RFC6960}.
Let's encrypt, for example, only provides revocation checks through OCSP.
Web servers can also staple the OCSP response in the TLS handshake if client explicitly requests it.

In addition, there are proprietary alternatives which aim to improve performance or reliability of CRL and OCSP.
Google, for example, maintains a list of revoked certificates (CRLSets\cite{Bingler2023}) partly by crawling CRL lists and uses this in Chromium-based browsers.
Microsoft\cite{Microsoft2023}, Mozilla\cite{Goodwin2015}, and Apple~\cite{Apple2022} also have similar revocation lists used in their products.
It should, however, be noted that such proprietary solutions have been show to be non-exhaustive.
Liu~\etal~\cite{Liu2015}, for instance, claims that CRLSets only cover 0.35\% of all revocations as observed by researchers.

\paragraph{Relying party policies}
Relying parties, \eg browsers, may also define additional validation policies.
For example, Apple products such as the Safari browser require at least 2 SCTs for certificates with a validity of less than 15 months~\cite{AppleCT2021}.

\subsection{User Perspective}
Users interact with TLS-secured server through a client: a web browser, mail client, smartphone app, or alike.
Depending on measurement approach, a discrepancy might arise between what collected data reveals and what users actually experience in their daily interactions over the Internet.
Here we show how users experience using a web browser might differ what measurements might reveal and discuss 4 aspects: \one point of access, \two redirects and resources, \three certificate validity, and \four bot and intrusion detection.

\paragraph{Point of access}
Domain names were introduced to take the burden of remembering IP addresses from users.
However, instead of remembering domain names and navigating to them, users navigate the Internet through the lens of search engines, social media, and streaming services~\cite{Domo2022}.
Research shows that users in part take browser address bar for a \emph{search bar} and have difficulties in differentiating components of domain names and URLs in general~\cite{Roberts2020}.

When choosing a data sources (see \autoref{sec:datasrc}), it should be taken into account that users might never (directly) encounter or interact with items in that source.
For example, \texttt{mtalk.google.com} \textemdash at position 110 of Cisco Umbrella Top 1M (for list \texttt{2023-06-21})\textemdash is an endpoint used by Google for push notifications and is never visited directly by users.
The same applies for IP addresses: when scanning the IP address space, endpoints such as home routers might be discovered that are only publicly accessible by misconfiguration.

\begin{figure*}
    \centering
    \scriptsize
    
    \begin{subfigure}[b]{.45\textwidth}
        \input{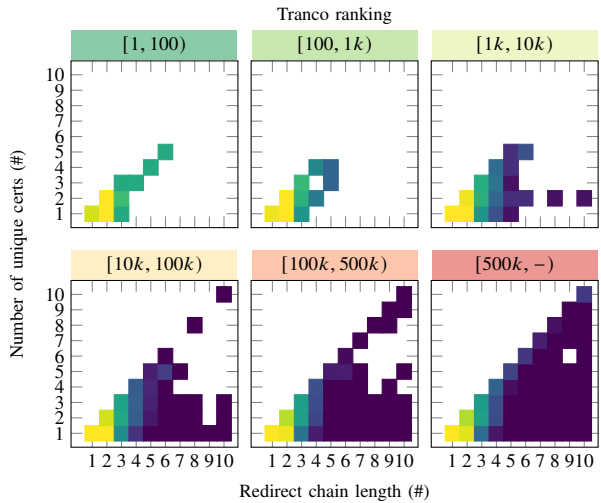}
        \caption{Redirect (\texttt{HTTP} and \texttt{HTML}) chains constructed by custom script based on \texttt{libcurl} (see \textbf{P.3} in \autoref{fig:toolchain}).}
        \label{subfig:red-curl}
    \end{subfigure}
    \hfill
    \begin{subfigure}[b]{.45\textwidth}
        \input{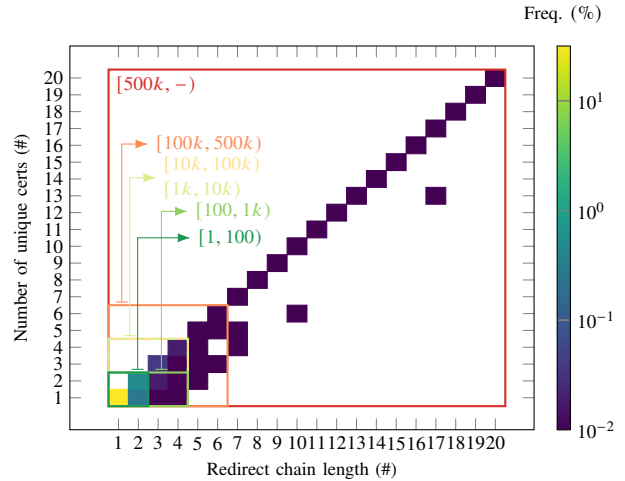}
        \caption{Redirect chains constructed by Puppeteer headless browser (see \textbf{M.1} in \autoref{fig:toolchain}). Only valid certificates are considered.}
        \label{subfig:red-browser}
    \end{subfigure}

    \caption{Frequency of unique certificates in redirect chain of a given domain name in relation to redirect chain length for hosts in our \texttt{TRANCO} dataset.}
    \label{fig:redirect}
\end{figure*}

\paragraph{Redirects and Resources}
Redirects are common on the Web: HTTP (\texttt{3xx} status codes), JavaScript (\texttt{window.location} property or History API), or HTML (\texttt{http-equiv} meta tag) all provide means to redirect a URL to another.
During redirects, browsers establish new TLS connections that are not signaled to the user.
For example, navigation to \texttt{google.com} triggers a redirect to \texttt{wwww.google.com} whereby only the connection and certificate information for the latter connection are presented to the user.
The connection attributes or certificates provided at each redirect are not necessarily the same as the previous or next step: \texttt{google.com} provides a different certificate compared to \texttt{wwww.google.com}.
Furthermore, when users navigate to a webpage, additional TLS connections to possibly different hosts are established to fetch resources such as scripts and stylesheets.

The role of redirects can be observed from the measurements of our \texttt{TRANCO} dataset as depicted in Figures \ref{subfig:red-curl} and \ref{subfig:red-browser}.
The former figure depicts the frequency of total \texttt{HTTP} and \texttt{HTML} redirects in relation to the total number of unique certificates as observed by our custom \texttt{libcurl} script (corresponding to step \textbf{P.3} in \autoref{fig:toolchain}).
Although the majority of domain names are redirected less than 3 times with 1 or 2 unique certificates in the redirect chain, we observe that as the ranking increases (\ie popularity decreases) both the length of redirect chain and number of unique certificates increase.
Note that certificates are not validated here.
The host names of TLS secured servers at the end of each redirect chain is then fed into a browser (corresponding to step \textbf{M.1} in \autoref{fig:toolchain}).
\autoref{subfig:red-browser} shows the results.
Here, JavaScript redirects are also followed while only valid certificates are considered.
Similarly, we observe that over $98\%$ of all hosts having no extra redirects.

\paragraph{Certificate validity}
Understanding how often users face invalid certificates, requires understanding the validation context as described above.
The impact of having the proper context is reflected in the discrepancy between studies that use generic contexts and those using the actual context of relying parties, \eg browsers.
Oakes~\etal~\cite{Oakes2019}, for example, collect complete certificates chain directly from users' machines
and observe that $99.7\%$ of all collected chains, \ie those used by clients, are valid.
In contrast, other studies based on IPv4 scans on generic root stores classify the majority of certificates (ranging between $66\%$ and $88\%$) as invalid~\cite{Chung2016,Farhan2023}.
Similarly, a study by Holz~\etal~\cite{Holz2016} observes that passively collected data exhibit salient percentages of valid certificates in comparison to certificates collected by active scanning of IPv4 address space.

Another source of discrepancy is the proprietary revocation lists discussed previously.
Liu~\etal~\cite{Liu2015}, for example, show how CRLSets (Google) only cover a fraction of all certificate revocations, so there is a chance that a Chromium-based browser would validate a certificate even though it has been revoked.
Note that none of the tools introduced in \autoref{sec:tooling} (excpet Puppeteer) have integrated nor apply these lists.

\begin{table}
    \scriptsize
    \caption{Difference between the length of valid certificate chains as delivered by servers in our \texttt{TRANCO} dataset and the length of the shortest valid chain built against Mozilla trust store (frequency  in \%).}
    \label{tab:chain-diff}
    \pgfplotstableset{
        my multistyler/.style 2 args={
        @my multistyler/.style={display columns/##1/.append style={#2}},
        @my multistyler/.list={#1}
    }}
    \pgfplotstableset{
        every head row/.style={
            before row={
            \toprule
            & \multicolumn{12}{c}{Length difference with the shortest verifiable chain} \\
            \cmidrule{2-13}
            },
            after row=\midrule,
        },
        every last row/.style={
            after row={
                \bottomrule
            }
        },
        after row=\dhline,
        create on use/Tranco ranking/.style={
            create col/set list={{$[1,100)$},{$[100,1k)$},{$[1k,10k)$},{$[10k,100k)$},{$[100k,500k)$},{$[500k,-)$}}},
        my multistyler={1,3}{sci sep align},
        my multistyler={2,4,5,6}{dec sep align},
        columns/Tranco ranking/.style={string type,column type = {l}},
        assign column name/.style={
            /pgfplots/table/column name={\textbf{#1}}
        },
    }

    \begin{filecontents*}{./figures/data/chain-diff-ranked-freq.csv}
rankGroup,-3,-2,-1,0,1,2
1-1c,0,0,0,69.2307692307692,25.6410256410256,5.12820512820513
1c-1k,0,0.138504155124654,0,73.1301939058172,16.7590027700831,9.97229916897507
1k-10k,0.01453065969195,0.30514385353095,0,68.9915722173787,20.3283929090381,10.3603603603604
10k-100k,0.00284742094847592,0.800125286521733,0.00142371047423796,65.5519013653384,25.2822506015177,8.36145161519953
100k-500k,0.00804484518499926,1.01493766853951,0.00193076284439982,67.1554714601072,26.7182180346121,5.10139722871173
500k-inf,0.00681353831648517,0.782590120553591,0.00082867357903198,69.7884028053363,26.7819934755767,2.63937138663791
\end{filecontents*}
\pgfplotstabletypeset[
        col sep=comma,
        columns={Tranco ranking,[index]1,[index]2,[index]3,[index]4,[index]5,[index]6},
        ]{./figures/data/chain-diff-ranked-freq.csv}
\end{table}

Finally, user clients are more robust in correcting malformed certificates chains (wrong order, duplicate certificates, missing intermediates, \etc) which might, for exmaple, cause OpenSSL to mark a certificate chain as invalid when using default settings.
In our measurement of \texttt{TRANCO} dataset, we observe that about 70\% of all hosts provide the shortest verifiable trust chain (from leaf to root) as depicted in \autoref{tab:chain-diff} (column $0$).
The rest is either providing more certificates (columns 1 and 2), \eg cross-signed variation of root certificates, or fewer certificates (columns -3, -2, and -1), \eg missing intermediate certificates.
Whereas providing extra certificates can be considered as an attempt to increase compatibility, \eg to cover cases where a root certificate is not yet included in a trust store, leaving intermediate certificates out might lead to certificate invalidation, \eg when intermediate certificates are not explicitly configured by the relying party.

\paragraph{Bot and Intrusion Detection}
Some infrastructure operators, \eg Akamai~\cite{Akamai2023}, offer \emph{bot detection} solutions to block unwanted or malicious access.
Such systems might block measurements if they flag the utilized tool as non-human and skew the results respectively.

\subsection{TLS security}
TLS Attacks make use of protocol flaws, implementation deficiencies, or cipher suite and extension weaknesses.
For an attack to be realistic, both endpoints must be susceptible to the respective attack vector.
For example, while security shortcomings of TLS~1.0 and~1.1 are known (see RFC8996~\cite{RFC8996}), none of mainstream browsers support these versions anymore, so that a web server supporting TLS versions 1.0 up to 1.3 does not in practice pose a higher security risk to users (with an up-to-date browser) than a server only supporting TLS~1.3.
Another example regards TLS session resumption: Cloudflare deploys a global memory cache to store TLS session parameters~\cite{Lin2015}.
It allows using a session ticket acquired for a given domain name (denoted in SNI extension) to be used for another domain name hosted at Cloudflare even if the certificate provided at the initial connection does not match the second domain name.
Browsers, however, bind TLS sessions to domain names and not just IP addresses, so that this specific situation cannot be abused without compromising the browser first.

\subsection{Data from Mixed Sources}
In \autoref{sec:prepare}, we discussed the impact of vantage point and importance of maintaining temporal integrity.
These aspects must also be regarded when interpreting data from mixed sources.
Cangialosi~\etal~\cite{Cangialosi2016}, for example, use WHOIS data from third party sources to distinguish between domain name owners and subsequently detect private key sharing among distinct service providers.
The temporal discrepancy between data from the original and the third party sources can lead to false identification of a domain name owner.
A study by Liu~\etal~\cite{Liu2015} is another example that relies on data from both passive and active measurements from different sources and times.

\subsection{CAA Resource Records}
DNS CAA records enable domain name owners to indicate which CAs are authorized to issue certificates for their names (see \autoref{sec:bg}).
Although CAA RRs are not relevant for users, from a research perspective they provide insights into internal organization of CAs.

The main challenge in verifying if a CAA records matches the certificate issuer is twofold: \one there is no centralized database containing a mapping between CAs and CAA value, and \two given a certificate, the actual issuing organization cannot always be uniquely detected as discussed above.
To address the former issue, Mozilla provides a list of CAAs as part of its Common CA Database~\footnote{\url{https://ccadb.my.salesforce-sites.com/ccadb/AllCAAIdentifiersReport}} which is composed on a best-effort basis and is neither exhaustive nor necessarily up-to-date.

For our dataset, we took Mozilla CAA identifiers from 2023-05-04 as a point of departure and manually matched each identifier with respective CA.
To do this, we looked at repositories of each CA and enumerated intermediate CAs.
The results reveal a rather confusing state of CAA identifiers that makes a mapping to CAs very difficult.
For example, DigiCert matches multiple CAA values (\eg \texttt{geotrust.com}, \texttt{thawte.com}, and \texttt{rapidssl.com}), and the Dutch Government relies on various CAs\footnote{\url{https://cert.pkioverheid.nl}} each with their own idenfiers (\eg \texttt{www.pkioverheid.nl} and \texttt{kpn.com}).

In our \texttt{TRANCO} dataset, we observe only 90,750 (3.76\%) of all entries provide an RFC conform CAA entry.
Out of which 15\% do not match the provided certificate using the method provided above, and about one third belong to Let's Encrypt certificates.
We also observe 1142 entries that are not parsable (wrong quoting or escaping, email addresses instead of domain names as value, \etc) or have the reserved bit set.

	\section{Conclusion}
\label{sec:conclusion}
The TLS protocol, as a mean to establish secure communication channels over the Internet, and the Web PKI, as a common basis for TLS authentication, have been a popular target of Internet measurements.
In this tutorial we first introduced a systematic approach to TLS, Web PKI, and their interplay, and used it to work out relevant aspect comprising potential data sources, measurement features and attributes, and temporal aspects of measurements.
We provided an overview of tools most commonly used for TLS and Web PKI measurements, discussed how to prepare, and perform measurements, and finally demonstrated various pitfalls in interpreting measurement data.

We discovered discrepancies and even contradictory findings in related works and discussed probable causes, and how such cases can be avoided when designing, performing, and interpreting own measurements.
In addition, the summary of related work of the past decade provided in this work can serve as an overview of which aspects of TLS and Web PKI measurements have previously been subject of study and how these measurements have been performed.

	\section*{Acknowledgments}
    This work was supported in parts by the German Federal Ministry of Education and Research (BMBF) within the projects PRIMEnet and IPv6Explorer.
	
	\balance
	\bibliographystyle{ieeetran}
	\bibliography{paper,rfc}

	\appendix
	\section*{Dataset Used in This Paper}
\label{sec:appendix}
\begin{figure*}
    \scriptsize
    \centering
    \input{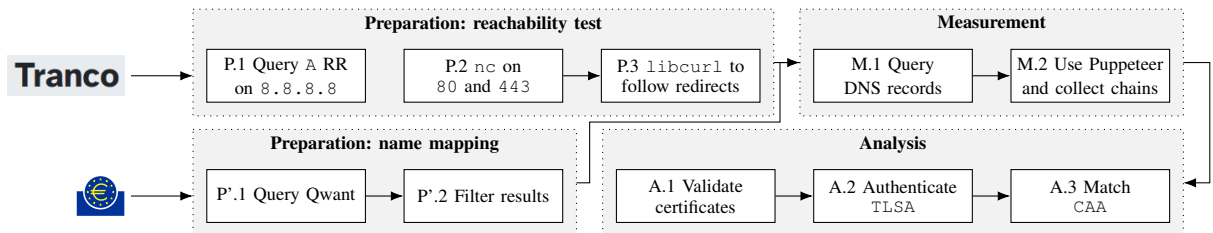}
    \caption{Simplified toolchain and procedure used in collection, measurement, and analysis of \texttt{TRANCO} and \texttt{EU-FINANCE} datasets}
    \label{fig:toolchain}
\end{figure*}

Beside the related work (summarized in \autoref{tab:survey}), we have conducted our own measurements which we use to exemplify and clarify the aspects that we discuss above.
The following comprises a brief overview of our dataset (see \autoref{tab:datasets}), and our measurement and analysis method (depicted in \autoref{fig:toolchain}).

\subsection{Dataset}
We use three sets, comprising domain or institution names, as point of departure for our measurements:

\begin{itemize}[itemindent=6em]
    \item[\texttt{TRANCO}] Tranco full list
    \item[\texttt{EU-FINANCE}] List of EU \emph{Monetary Financial Institutions}
    \item[\texttt{IR-BANKS}] List of Iranian Banks
\end{itemize}

\begin{table}
    \scriptsize
    \caption{Datasets used in this paper}
    \label{tab:datasets}
    \begin{tabularx}{\columnwidth}{>{\tt}r r X r r}
        \toprule
        \textnormal{Name} & Entries & Source & Access Date & Measurement Date\\
        \midrule
        TRANCO & 3.77~M & \url{tranco-list.eu} & 2023-06-14 & 2023-11-07 \\
        \dhline
        EU-FINANCE & 4~k & EU Central Bank & 2023-06-12 & 2023-06-\{12,14\} \\
        \dhline
        IR-BANKS & 28 & Online search & 2023-06-15 & 2023-06-15 \\
        \bottomrule
    \end{tabularx}
\end{table}

\noindent
Each dataset has been selected due to its unique properties.
The first dataset, \texttt{TRANCO}, comprises most popular domain names (see \autoref{sec:datasrc}).
The second dataset, \texttt{EU-FINANCE}, was selected under the assumption that financial institutions apply security precautions that might be reflected in their choice of certificates or TLS setup.
Finally, the list of Iranian banks, \texttt{IR-BANKS}, represents a group of institutions under US and EU sanctions but reliant on services from those country for their online operation.
\autoref{tab:datasets} sumarizes these datasets.

\subsection{Method}
We perform our measurements from a single vantage point in Germany.
Before performing actual measurements, we run our initial data through two preparation pipelines.
For the \texttt{TRANCO} dataset, we perform a reachability test and discard any domain name that doesn't resolve to an IP address or is not reachable through ports \texttt{80} or \texttt{443}.
In case of \texttt{EU-FINANCE} dataset, preparation phase maps real-world institution names to domain names.
\autoref{fig:toolchain} depicts our method.
Note that none of the entries in \texttt{IR-BANKS} are reachable from outside Iran so that our toolchain is not applicable.
For this set we limit ourselves to X.509 certificates that we manually collected from CT Logs.

\paragraph{Reachability test}
Domain names in the \texttt{TRANCO} dataset are not necessarily delegated or resolve to an IP address (see \autoref{sec:datasrc}).
To filter out unreachable hosts, we first try to resolve the given name to an IPv4 address (using Google \texttt{8.8.8.8} recursive resolvers).
$12.8\%$ of entries do not resolve to an IP address and a timeout occurs in 249 cases.

For all reachable hosts, we scan to see if ports \texttt{80} and \texttt{443} (\texttt{HTTP} and \texttt{HTTPS} respectively) are open.
About $12 k$ does not have port \texttt{80} open, $348 k$ port \texttt{443}, and $173 k$ have neither ports open.

For all hosts and open ports we use a custom tool based on \texttt{libcurl} to contact the host over \texttt{HTTP} or \texttt{HTTPS}.
In this process, we also follow \texttt{HTTP} and \texttt{HTML} redirects and record all intermediate domain names.
We end up with $2.5 M$ host names that were accessible over \texttt{HTTPS}.
The host names from all \texttt{HTTPS} URLs at the end of each redirect chain are then used for the actual measurement as discussed below.
Note that we don't validate certificates at this step.

\paragraph{Name mapping}
For our \texttt{EU-FINANCE}, we first need to map real world names to domain names.
For this, we use Qwant, a French search engine which provides a free and publicly accessible API.
For each entry, we query the institution name and store matching results.
We then generate a list of all domain names over all queries and sort it by occurrence frequency.
Based on this list, we generate a blacklist of social media, encyclopedia, yellow pages, and alike.
For each result page per institution we remove entries matching the blacklist and take the first remaining entry as the matching domain.
Finally, the results are manually verified to remove any possible false positives.
It should be noted that in contrast to entries in \texttt{TRANCO} dataset, we already assume reachability for the results of this phase (search bot visits).

\paragraph{Measurement}
For the actual measurement, we feed processed domain names from \texttt{TRANCO} and \texttt{EU-FINANCE} datasets to Puppeteer (a headless browser; see \autoref{sec:tooling}).
Puppeteer enables us to have a similar experience to users when navigating to these domains.
For each domain name, we again follow redirects (including JavaScript redirects) and store X.509 certificate chains alongside \texttt{HTTP} headers.
We also collect \texttt{A}, \texttt{SOA}, \texttt{NS}, \texttt{TLSA}, and \texttt{CAA} DNS records for each domain name.

\paragraph{Data analysis}
We use the crypto library from ZMap project to parse and validate certificate chains (step \textbf{A.1}) against Mozilla Root Store from 2023-01-10\footnote{\url{https://curl.se/docs/caextract.html}}.
To imitate browser behavior, we also manually include all non-expired and non-revoked intermediate certificates (from 2023-01-21\footnote{\url{https://wiki.mozilla.org/CA/Intermediate_Certificates}}) chaining up to the root certificates described above.
In cases where multiple validation paths were available, for example, due to cross certification, we used the shortest for our analysis, \eg CA market share.
For DANE authentication we used the library from Shumon Huque\footnote{\url{https://github.com/shuque/dane}}.
To match CAA records with leaf certificate issuers we took list of CAA identifiers from Mozilla Common CA Database, compared them with PKI repositories of respective CAs and enhanced the list with our own data.

\end{document}